\documentclass[sigconf]{acmart}
\PassOptionsToPackage{bookmarks={false}}{hyperref}
\hypersetup{draft}
\usepackage{xcolor}
\usepackage{booktabs} 
\usepackage[ruled,vlined,linesnumbered]{algorithm2e}
\SetAlFnt{\small}
\SetAlCapFnt{\small}
\SetAlCapNameFnt{\small}
\SetAlCapHSkip{0pt}
\IncMargin{-\parindent}

\usepackage{balance}
\usepackage[inline, shortlabels]{enumitem}

\usepackage{subfig} 
\usepackage{wrapfig}
\usepackage{listings}
\usepackage{multicol}
\usepackage{color, colortbl}
\colorlet{lightgreen}{green!25}
\colorlet{lightblack}{yellow!50}
\colorlet{lightblue}{orange!50}
\colorlet{lightred}{red!50}

\definecolor{lightgray}{gray}{0.97}
\definecolor{codeblue}{rgb}{0.13,0.13,1}
\definecolor{OliveGreen}{cmyk}{0.64,0,0.95,0.40}
\definecolor{DarkGreen}{cmyk}{0.40,0,0.50,0.15}
\definecolor{purple}{cmyk}{0.41,0.73,0,0}
\definecolor{LightRed}{cmyk}{0,0.682,0.728,0}

\usepackage{pifont}
\usepackage{graphicx}
\usepackage[skip=5pt]{caption}

\newcommand{\para}[1]{\vspace{0.1em}{\bf #1.}\ }

\newcommand{\cmark}{\ding{51}}%
\newcommand{\xmark}{\ding{55}}%

\newcommand{\figref}[1]{Figure~\ref{#1}}
\newcommand{\tabref}[1]{Table~\ref{#1}}
\newcommand{\fignref}[1]{Figure~\ref{#1}}
\newcommand{\algoref}[1]{Algorithm~\ref{#1}}

\newcommand{\secref}[1]{\S\ref{#1}}
\newcommand{\secnref}[1]{\S\ref{#1}}

\newcommand{\footnoteref}[1]{$^{\ref{#1}}$}
\newcommand{\etal}{~\textit{et al.}}
\newcommand{\graphprop}{cyclicity}
\newcommand{\Graphprop}{Cyclicity}

\newcommand{\name}{bespoke control flow analysis\xspace}
\newcommand{\shortname}{BCFA\xspace}

\copyrightyear{2020}
\acmYear{2020}
\setcopyright{acmcopyright}
\acmConference[ICSE'20]{42nd International Conference on Software Engineering}{May 23--29, 2020}{Seoul, Republic of Korea}
\acmBooktitle{42nd International Conference on Software Engineering (ICSE '20), May 23--29, 2020, Seoul, Republic of Korea}
\acmPrice{15.00} 
\acmDOI{10.1145/3377811.3380435} 
\acmISBN{978-1-4503-7121-6/20/05}

\begin{document}
\title{BCFA: Bespoke Control Flow Analysis for CFA at Scale}

\author{Ramanathan Ramu$^1$}
\thanks{$^1$~At the time this work was completed, Ramanathan Ramu 
was a graduate student at Iowa State University. 
$^2$~At the time this work was completed, Ganesha Upadhyaya 
was a graduate student at Iowa State University.
$^3$~At the time this work was completed, Dr. Hoan Nguyen was a 
postdoctoral researcher at Iowa State University.}
\affiliation{
	\institution{Microsoft}            
	\city{Redmond}
  	\state{Washington}
}
\email{ramu.rm2013@gmail.com}

\author{Ganesha B Upadhyaya$^2$}
\affiliation{
	\institution{Harmony}            
	\city{Mountain View}
  	\state{California}
}
\email{ganeshau@iastate.edu}

\author{Hoan Anh Nguyen$^3$}
\affiliation{
	\institution{Amazon}            
	\city{Seattle}
  	\state{Washinton}
}
\email{nguyenanhhoan@gmail.com}

\author{Hridesh Rajan}
\affiliation{
	\institution{Dept. of Computer Science, Iowa State University}            
	\city{Ames}
  	\state{Iowa}
}
\email{hridesh@iastate.edu}

\renewcommand{\shortauthors}{R. Ramu et al.}

\begin{abstract}
Many data-driven software engineering tasks such as discovering programming
patterns, mining API specifications, etc., perform source code analysis over
control flow graphs (CFGs) at scale. Analyzing millions of CFGs can be expensive
and performance of the analysis heavily depends on the underlying CFG traversal
strategy. State-of-the-art analysis frameworks use a fixed traversal strategy.
We argue that a single traversal strategy does not fit all kinds of analyses and
CFGs and propose {\em \name} (\shortname). Given a control flow analysis (CFA)
and a large number of CFGs, \shortname selects the most efficient traversal
strategy for each CFG.
\shortname extracts a set of properties of the CFA by analyzing the code of the
CFA and combines it with properties of the CFG, such as branching factor and
cyclicity, for selecting the optimal traversal strategy.
We have implemented \shortname in {\em Boa}, and evaluated \shortname using a
set of representative static analyses that mainly involve traversing CFGs and
two large datasets containing 287 thousand and 162 million CFGs. Our results
show that \shortname can speedup the large scale analyses by 1\%-28\%. Further,
\shortname has low overheads; less than 0.2\%, and low misprediction rate; less
than 0.01\%.
%

\end{abstract}

%


%

\maketitle

\section{Introduction}
\label{sec:introduction}
Data-driven techniques have been increasigly adopted in many software
engineering (SE) tasks: API precondition mining~\cite{preconditions,khairunnesa2017}, API usage
mining~\cite{acharya-etal,zhang-etal}, code search~\cite{mcMillan-etal},
discovering vulnerabilities~\cite{yamaguchi-etal}, to name a few. These
data-driven SE tasks perform source code analysis over different program
representations like source code text, abstract syntax trees (ASTs),
control-flow graphs (CFGs), etc., at scale. For example, API precondition mining
analyzes millions of methods that contain API call sites to capture conditions
that are often checked before invoking the API. The source code mining
infrastructures~\cite{dyer2013boa,Bajracharya-etal-04,Gousi13} have started
supporting CFG-level analysis to facilitate variety of data-driven SE tasks.


Performance of source code analysis over CFGs heavily depends on the order of
the nodes visited during the traversals: {\em the traversal strategy}.
Several graph traversal strategies exist from the graph traversal literatures,
e.g., depth-first, post-order, reverse post-order, topological order,
worklist-based strategies, etc. However, the state-of-the-art analysis frameworks
use fixed traversal strategy.
For example, \texttt{Soot} analysis framework~\cite{soot} uses topological
ordering of the nodes to perform control flow analysis. Our observation is that
for analyzing millions of programs with different
characteristics, no single strategy performs best for all kinds of analyses and
programs. Both properties of the analysis and the properties of the input
program influence the traversal strategy selection. For example, for a control flow
analysis that is data-flow sensitive, meaning the output for any node must be
computed using the outputs of its neighbors, a traversal strategy that visits
neighbors prior to visiting the node performs better than other kinds of
traversal strategies. Similarly, if the CFG of the input program is sequential,
a simple strategy that visits nodes in the random order performs better than a
more sophisticated strategy.

We propose {\em \name} (\shortname), a novel source code analysis technique for
performing large scale source code analysis over the control flow graphs.
Given an analysis and a large collection of
CFGs on which the analysis needs to be performed, \shortname selects an optimal
traversal strategy for each CFG.
In order to achieve that, \shortname deploys a novel
decision tree that combines a set of analysis properties with a set of graph
properties of the CFG. The analysis properties include data-flow sensitivity,
loop sensitivity, and traversal direction, and the graph properties include
cyclicity (whether the CFG contains branches and loops). There exists no
technique that automatically selects a suitable strategy based on analysis and
CFG properties. Since manually extracting the properties can be infeasible when
analyzing millions of CFGs, we provide a technique to extract the analysis
properties by analyzing the source code of the analysis.

We have implemented \shortname in Boa, a source code mining
infrastructure~\cite{dyer2013boa,boa-b} and evaluated \shortname using a set of 21 source code
analyses that includes mainly control and data-flow analyses.
The evaluation is performed on two datasets: a dataset containing
well-maintained projects from DaCapo benchmark (with a total of 287K CFGs),
and an ultra-large dataset containing more than 380K projects from GitHub
(with a total of 162M CFGs).
Our results showed that \shortname can speedup the large scale analyses by
1\%-28\% by selecting the most time-efficient traversal strategy.
We also found that, \shortname has low overheads for computing the analysis and
graph properties; less than 0.2\%, and low misprediction rate; less than 0.01\%.

In summary, this paper makes the following contributions:
\begin{itemize}[leftmargin=*]
  \item It proposes a set of analysis properties and a set of graph properties
  that influence the selection of traversal strategy for CFGs.
  \item It describes a novel decision tree for selecting the most suitable traversal 
  strategy using the analysis and the graph properties. 
  \item It provides a technique to automatically extract the analysis properties
  by analyzing the source code of the analysis. 
  \item It provides an implementation of \shortname in Boa~\cite{dyer2013boa,boa-b}
  and a detailed evaluation using a wide-variety of source code analyses
  and two large datasets containing 287 thousand and 162 million CFGs.
\end{itemize}
\section{Motivation}
\label{sec:motivation}
Consider a software engineering task that infers the temporal specifications
between pairs of API method calls, i.e., a call to $a$ must be followed by a
call to
$b$~\cite{engler-sosp2001,ramanathan-icse2007,weimer-tacas2005,yang-icse2006}.
A data-driven approach for inference is to look for pairs of API calls that
frequently go in pairs in the same order at API call sites in the client code.
Such an approach contains (at least) three source code analyses on the control
flow graph (CFG) of each client method:
1) identifying references of the API classes and call sites of the API methods
which can be done using \emph{reaching definition}
analysis~\cite{programanalysis};
2) identifying the pairs of API calls ($a$, $b$) where $b$ follows $a$ in the
client code using \emph{post-dominator} analysis~\cite{compilers}; and
3) collecting pairs of temporal API calls by traversing all nodes in the
CFG---let us call this \emph{collector} analysis.
These analyses need to be run on a large number of client methods to produce
temporal specifications with high confidence.

Implementing each of these analyses involves traversing the CFG of each client
method. The traversal strategy could be chosen from a set of standard strategies
e.g., depth-first search (DFS), post-order (PO), reverse post-order (RPO),
worklist with post-ordering (WPO), worklist with reverse post-ordering (WRPO)
and any order (ANY).\footnote{In DFS, successors of a node are visited prior to
visiting the node. PO is similar to DFS, except that there is no order between
the multiple successors. In RPO, predecessors of a node are visited prior to
visiting the node. WPO and WRPO are the worklist-based strategies, in which the
worklist is a datastructure used to keep track of the nodes to be visited,
worklist is initialized using either PO or RPO (PO for WPO and RPO for WRPO),
and whenever a node from the worklist is removed and visited, all its
successors (for forward direction analysis) or its predecessors (for backward
direction analysis) are added to the worklist.}

\setlength{\belowcaptionskip}{-5pt}
\begin{figure}[t]
\centering
\includegraphics[width=\linewidth]{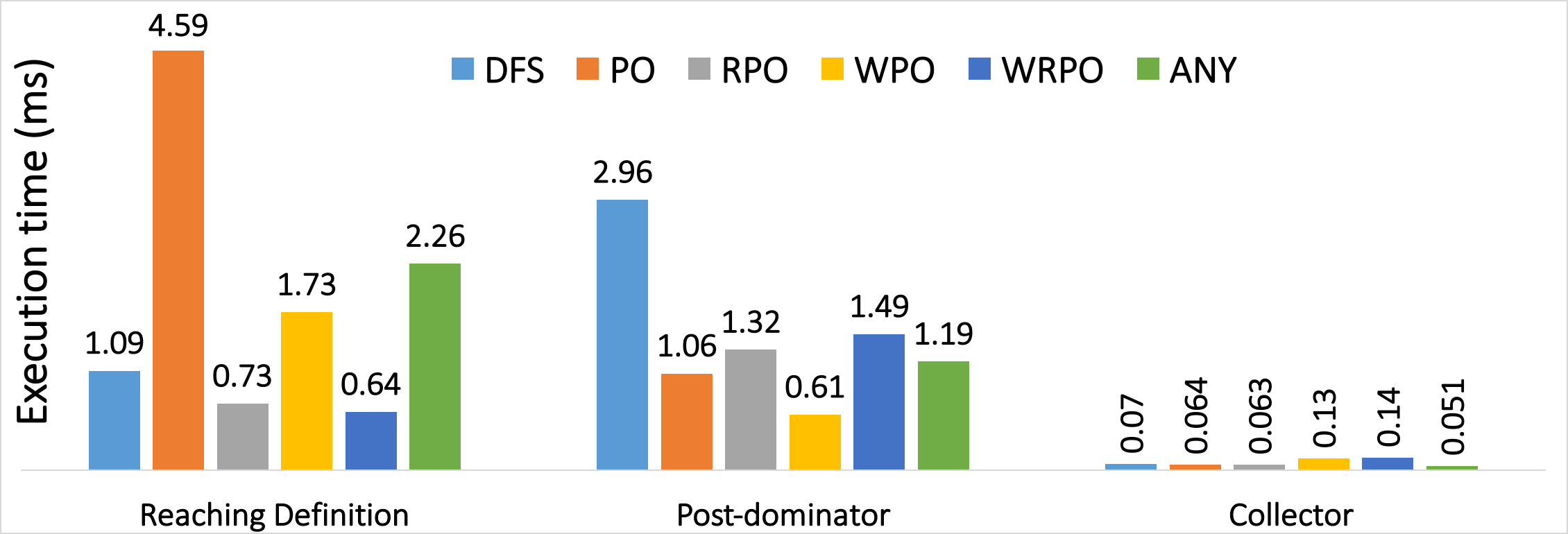}
\caption{Running times (in ms) of the three analyses on graph A using different
traversal strategies (lower better).}
\label{fig:example-a-graph}
\end{figure}

Figure \ref{fig:example-a-graph} shows the performance of these three analyses
when using standard traversal strategies on the CFG, referred to as A, of a
method in the DaCapo benchmark~\cite{blackburn2006dacapo}.
The CFG has 50 nodes and it has branches but no loops. Figure
\ref{fig:example-a-graph} shows that, for graph A, WRPO performs better than
other strategies for the reaching definition analysis, while WPO outperforms the
others for the post-dominator analysis and the ANY traversal works best for the
collector analysis.

{\bf No Traversal Strategy Fits All and Analysis Properties Influence the
Selection.\ }
The performance results show that there is no single traversal strategy that
works best for all analyses.
Reaching definition analysis is a forward data-flow analysis where the output at
each node in the graph is dependent on the outputs of their predecessor nodes.
So, DFS, RPO and WRPO by nature are the most suitable.
However, worklist is the most efficient strategy here because it visits only the
nodes that are yet to reach fixpoint\footnote{\label{fixpoint}Fixpoint is a
state where the outputs of nodes no longer change.} unlike other strategies
that also visit nodes that have already reached fixpoint.
Post-dominator analysis, on the other hand, is a backward analysis meaning that
the output at each node in the graph is dependent on the outputs of their
successor nodes. Therefore, the worklist with post-ordering is the most
efficient traversal.
For the collector analysis, any order traversal works better than other
traversal strategies for graph A. This is because for this analysis the output
at each node is not dependent on the output of any other nodes and hence it is
independent of the order of nodes visited.
The any order traversal strategy does not have the overhead of visiting the
nodes in any particular order like DFS, PO, RPO nor the overhead to maintain the
worklist.

{\bf Properties of Input Graph Determine Strategy.\ }
For the illustrative example discussed above, DFS and RPO were worse than WRPO
for the reaching definition analysis and PO was worse than WPO for
post-dominator because they require one extra iteration of analysis before
realizing that fixpoint has been reached.
However, since graph A does not have any loops, if the graph A's nodes are
visited in such a way that each node is visited after its predecessors for
reaching definition analysis and after its successors for post-dominator
analysis, then the additional iteration is actually redundant. Given that graph
A has no loops, one could optimize RPO or PO to skip the extra iteration and
fixpoint checking. Thus, the optimized RPO or PO would run the same number of
iterations as the respective worklist-based ones and finish faster than them
because the overhead of maintaining the worklist is eliminated.

\begin{figure}
\centering
\includegraphics[width=0.6\linewidth]{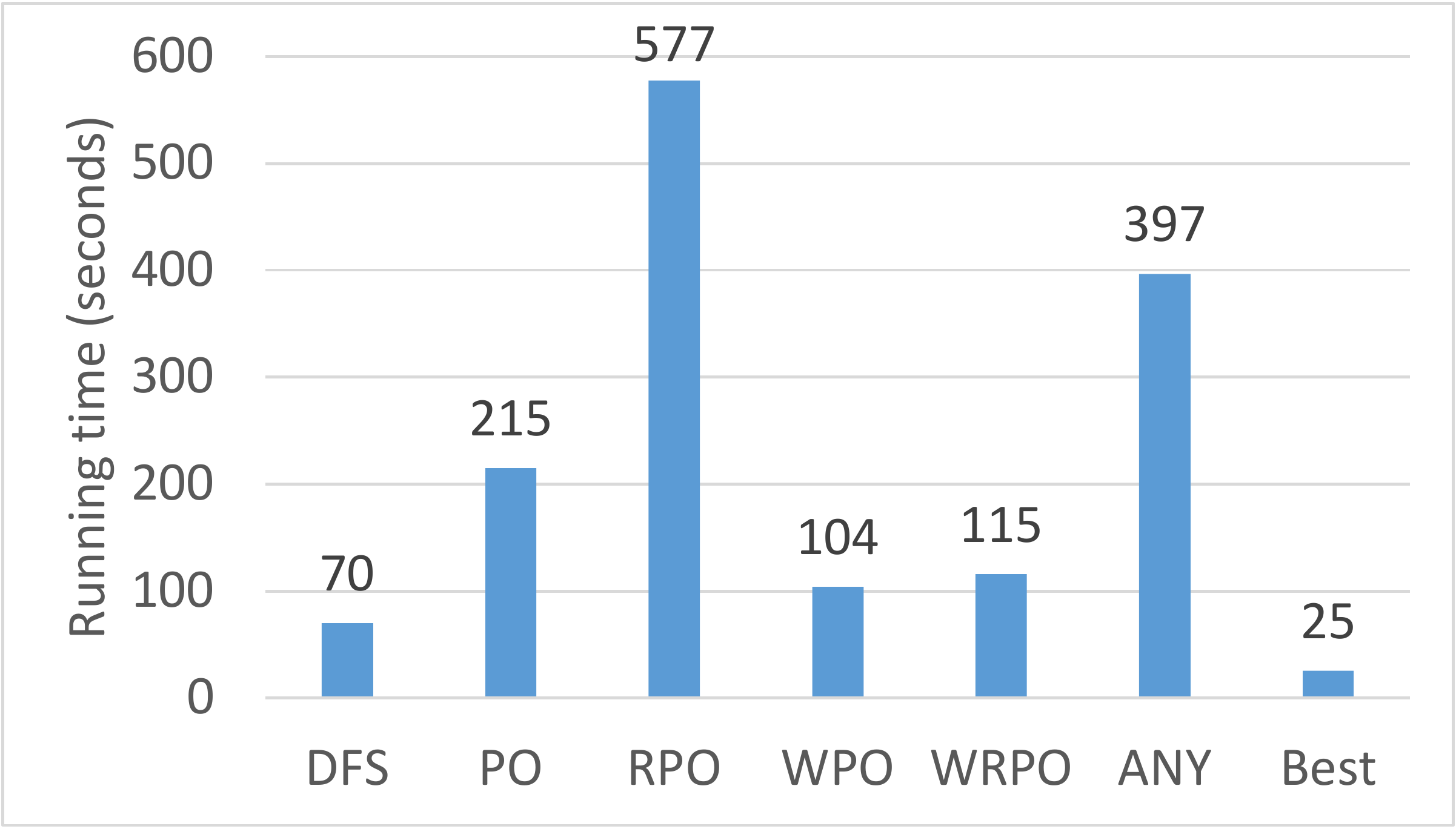}
\caption{Running times (in s) of three analyses using different traversal
strategies on a large codebase (lower better). }
\label{fig:motive}
\end{figure}

The potential gains of selecting a suitable traversal strategy could be
significant.
To illustrate, consider \figref{fig:motive} that shows the performance of our
entire illustrative example (inferring temporal specifications) on a large
corpus of 287,000 CFGs extracted from the DaCapo benchmark
dataset~\cite{blackburn2006dacapo}.
The bar chart shows the total running times of the three analyses. \textbf{Best}
is an ideal strategy where we can always choose the most efficient with all
necessary optimizations. The bar chart confirms that \textbf{Best} strategy
could reduce the running time on a large dataset from 64\% (against DFS) to 96\%
(against RPO). Three tasks could be completed for the price of one. 

On the other hand, the poor selection of the traversal strategy could make an
analysis infeasible when performed at scale. Consider
application~\cite{preconditions,khairunnesa2017} that uses post-dominator (PDOM) and reaching
definitions (RD) analyses. For PDOM, a poor strategy (e.g., RPO,
WRPO) would make the analysis never finish on our GitHub dataset.
Similarly, for RD, a poor strategy (WPO on GitHub) would be ~80\% slower.
\fignref{tab:reduction-analyses} shows several `-` for the GitHub column,
indicating that picking a poor strategy makes the analysis infeasible.

\section{\name}
\label{sec:approach}
\label{sec:overview}
 
\begin{figure}[ht!]
  \centering \includegraphics[width=\linewidth]{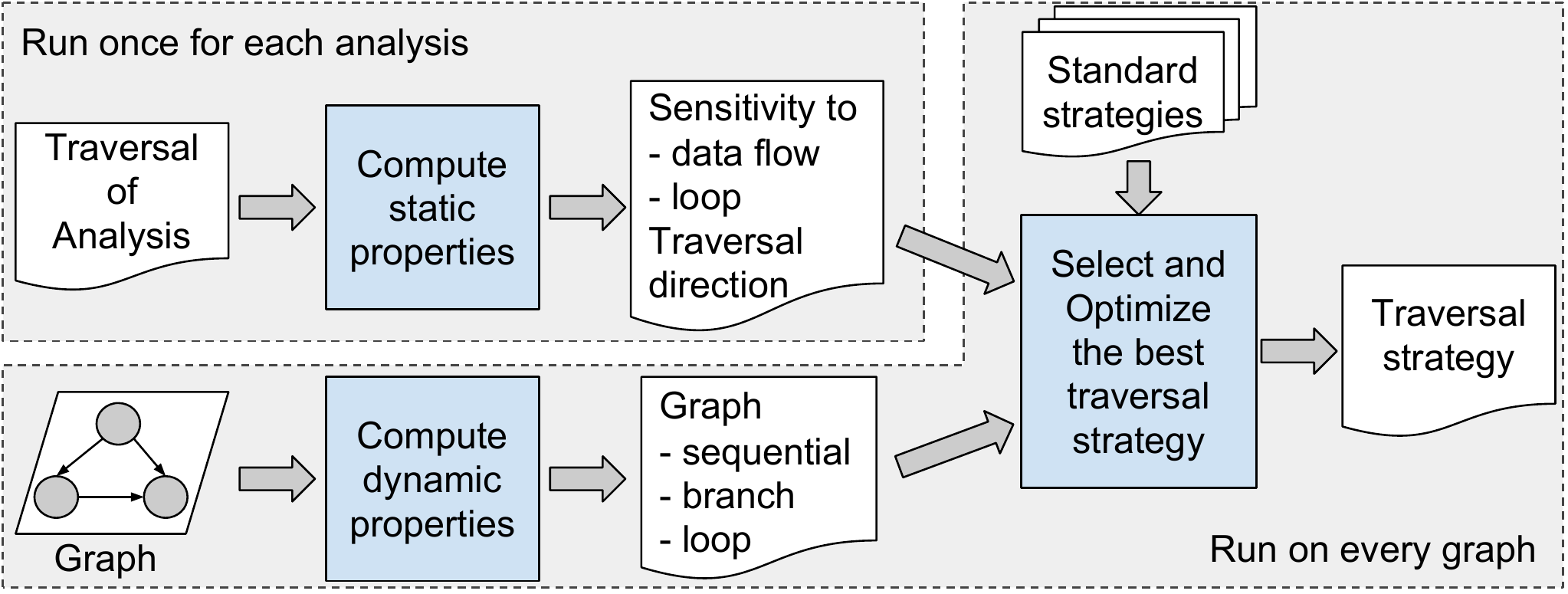}
\caption{Overview of \name.}
  \label{fig:overview}
\end{figure}

\fignref{fig:overview} provides an overview of \shortname and its key
components.
The inputs are a source code analysis that contains one or more
traversals (\secref{sec:language}), and a graph.
The output is an optimal traversal strategy for every traversal
in the analysis. For selecting an optimal strategy for a traversal, \shortname
computes a set of static properties of the analysis
(\secref{sec:compute-properties}), such as data-flow and loop sensitivity, and
extracts a runtime property about the \graphprop{} in the graph
(sequential/branch/loop). Using the computed analysis and graph properties, \shortname
selects a traversal strategy from a set of candidates
(\secref{sec:candidates}) for each traversal in the analysis
(\secref{sec:decision-tree}) and optimizes it (\secref{subsec:optimizations}).
The analysis properties are computed only once for each analysis, whereas graph
\graphprop{} is determined for every input graph.

\subsection{The Source Code Analysis Model}
\label{subsec:analysis-model}
A source code analysis can be performed on either the source code text or the
intermediate representations like control flow graph (CFG), program dependence
graph (PDG), and call graph (CG). An {\em iterative data-flow analysis} over a
CFG traverses the CFG by visiting the nodes and computing outputs for nodes (aka
analysis facts)~\cite{atkinson2001implementation}. For computing an output for a
node, the analysis may use the outputs of node's neighbors (successors or
predecessors). For example, consider the {\em Reaching definitions (RD)}
analysis~\cite{programanalysis} that computes a set of variable definitions that
are available at each node in a CFG. Here, the output or the analysis fact at a
node is a set of variables. The \texttt{RD} analysis performs two traversals of
the CFG, where in the first traversal it visits every node in the CFG starting
from the entry node and collects variables defined at nodes, and in the second
traversal, the variables defined at the predecessors are propagated to compute
the final set of variables at nodes. Note that, certain nodes may be visited
several times (in case of loops in the CFG) until a fixpoint is
reached.\footnoteref{fixpoint} A complex analysis may further contain one or
more sub-analyses that traverses the CFG to compute different analysis facts.

\subsection{Static Properties of the Input Analysis}
\label{sec:compute-properties}
These properties are the data-flow sensitivity, loop
sensitivity, and traversal direction. We now describe these properties in
detail and provide a technique to automatically compute them in
\secnref{sec:language}.
%

\subsubsection{Data-Flow Sensitivity}
\label{sec:data-flow}
The {\em data-flow sensitivity} property of a traversal models the dependence
among the traversal outputs of nodes in the input graph. A traversal is
data-flow sensitive if the output of a node is computed using the outputs of
other nodes. For example, the {\em Reaching definitions (RD)} analysis described
previously performs a traversal that aggregates the variable definitions from
the predecessors of a node while computing the output for the node, hence it is
considered data-flow sensitive.

A key observation that can be made from the description of the data-flow
sensitivity property is that, a traversal strategy that visits the neighbors of
a node prior to visiting the node will be more efficient than other kinds of
traversal strategies, because the output of neighbors will be available prior to
computing the output for the node and hence fixpoint can be reached faster.

\subsubsection{Loop Sensitivity}
The {\em loop sensitivity} property models the effect of loops in the input
graph on the analysis. If an input graph contains loops and if the
analysis traversal is affected by them, the traversal may require multiple
iterations (where certain nodes are visited multiple times) to compute the final
output at nodes. In these iterations, the traversal outputs at nodes either
shrink or expand to reach a fixpoint.\footnote{Shrink and expand are defined
with repsect to the size of the datastructure used to store the analysis outputs
at nodes. As the size is a dynamic factor, we determine the shrink or
expand property by tracking the operations performed on the datastructure used
to store the output. For example, ADD operation always expands and REMOVE always
shrinks. \secnref{sec:language} describes this in detail.} Hence, a traversal is
loop sensitive if the output of any node shrinks or expands in the subsequent
iterations. As one could imagine the loop-sensitivity property can be difficult
to compute, hence we have provided a technique in \secnref{sec:language}. The
{\em Reaching definitions (RD)} analysis described previously has two traversal,
in which the first traversal that simply extracts the variables defined at nodes
is not loop-sensitive, whereas the second traversal that propagates the
variables from the predecessors is loop-sensitive.
{\em If an analysis traversal is not loop-sensitive, the loops in the input CFG
will not have any influence on the selection of the traversal strategy.}

\subsubsection{Traversal Direction}
A {\em traversal direction} is property that describes the direction of visiting
the CFG. A CFG can be visited in either the \texttt{FORWARD} direction or the
\texttt{BACKWARD} direction. In \texttt{FORWARD}, the entry node
of the CFG is visited first, followed by the successors of the nodes, where in
\texttt{BACKWARD}, the exit nodes of the CFG are visited first, followed by the
predecessors of the nodes. The traversal direction is often specified by the
analysis writers.

\subsection{Graph \Graphprop~ of Input CFGs}
So far we have described the three static properties of the analysis that
influences the traversal strategy selection. A property, \graphprop{}, 
of the input graph also influences the selection. 
Based on the \graphprop{}, we classify graphs into four categories:
sequential, branch only, loop w/o branch and loop w/ branch. In
sequential graphs, all nodes have no more than one successor and
predecessor. In graphs with branches, nodes may have more than one
successors and predecessors. In graphs with loops, there exist cycles in
the graphs. The graph \graphprop{} is determined during graph construction. 

Graph cyclicity plays an important role in the selection of the appropriate
traversal strategy along with the analysis property for an (analysis, CFG) pair.
For CFGs with branches and loops, the outputs of all dependent nodes of a
node (predecessors or successors) may not be available at the time of visiting
the node, hence a traversal strategy must be selected that guarantees that the
outputs of all dependent nodes of a node are available prior to computing the
node's output, as it leads to fixpoint convergence in fewer iterations.

\subsection{Candidate Traversal Strategies}
\label{sec:candidates}
For every (analysis, CFG) pair, \shortname selects a traversal strategy from
among the candidate traversal strategies listed next.
\textit{1) Any order (ANY):} nodes can be visited in any order
(random). 
\textit{2) Increasing order of node ids (INC):} nodes are visited in
the increasing order of node ids. Node ids are assigned during the CFG
construction based on the control flow order.
\textit{3) Decreasing order of node ids (DEC):} nodes are visited in the
decreasing order of node ids.
\textit{4) Postorder (PO):} successors of a node is visited before visiting the
node.
\textit{5) Reverse postorder (RPO):} predecessors of a node is visited before
visiting the node.
\textit{6) Worklist with postorder (WPO):} nodes are visited in the order they
appear in the worklist. Worklist is a datastructure used to keep track of the
nodes to be visited. Worklist is initialized with postordering of the
nodes. Whenever a node from the worklist is removed and visited, all its
successors (for FORWARD) or predecessors (for BACKWARD) are added to the
worklist as described in \cite{atkinson2001implementation}.
\textit{7) Worklist with reverse postorder (WRPO):} similar to WPO, except that
the worklist is initialized with reverse postordering. 

The traversal strategies were selected by carefully reviewing the compiler
textbooks, implementations, and source code analysis frameworks. The selected
strategies are generally applicable to any analyses and graphs.\footnote{We did
not include strategies like chaotic iteration based on weak topological ordering
because they are effective only for computing fixpoint of a continuous function
over lattices of infinite height~\cite{bourdoncle93}.}

\subsection{Traversal Strategy Decision Tree}
\label{sec:decision-tree}
\begin{figure}[ht!]
\centering
\includegraphics[width=\linewidth]{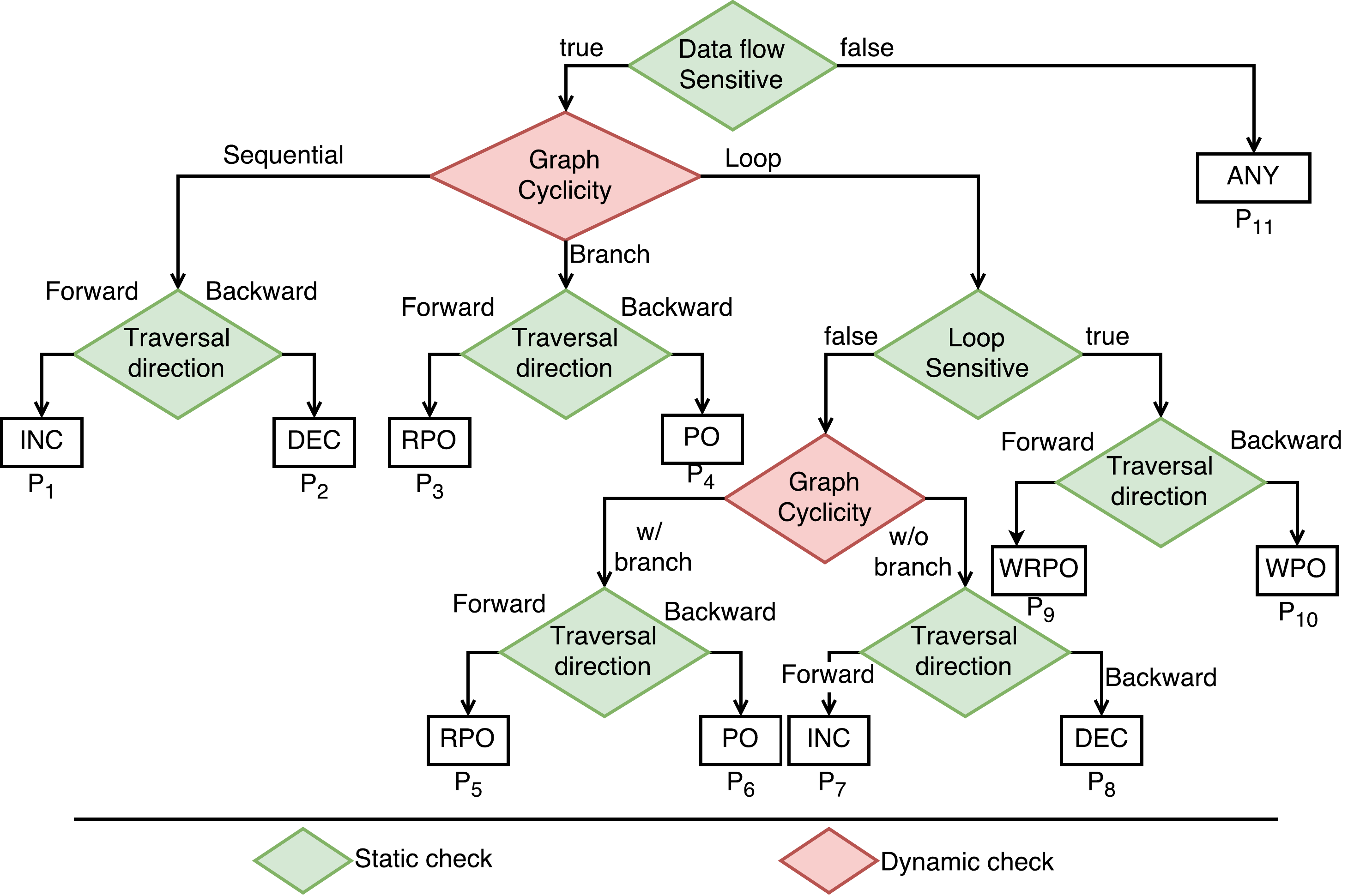}
\caption{\shortname decision tree for selecting an optimal traversal strategy. $P_0$ to
$P_{11}$ are markings for paths.}
\label{fig:decision-diagram}
\end{figure}

The core of \shortname is a 
decision tree for selecting an optimal traversal strategy shown in
\fignref{fig:decision-diagram}. 
The leaf nodes of the tree are one of the seven traversal strategies and
non-leaf nodes are the static and runtime checks. The decision tree has eleven
paths marked from $P_1$ through $P_{11}$. Given a traversal and an input graph,
one of the eleven paths will be taken to decide the best traversal strategy. The static
checks are marked green and the runtime checks are marked red. The static
properties includes
data-flow sensitivity ($P_{DataFlow}$), loop sensitivity ($P_{Loop}$), and
traversal direction. The runtime property that is checked is the graph
\graphprop{}: sequential, branch, loop w/ branch, and loop w/o branch. 
Next, we describe the design rationale of the decision tree.
To illustrate, we use the post-dominator analysis
described in \autoref{lst:dominators}, which has one data-flow insensitive
traversal \lstinline|initT| and one data-flow sensitive traversal
\lstinline|domT|, on a CFG shown in \autoref{fig:running-example} that has both
branches and loops.

The first property is the data-flow sensitivity (\secref{sec:data-flow}) which 
is computed by~\algoref{algo:flow-sensitive}. 
The rationale for checking this property first is
that, if a traversal is not data-flow sensitive, 
the traversal can finish in a single iteration (no fixpoint computation is 
necessary), irrespective of the type of the input graph. In
such cases, visiting nodes in any order will be efficient, hence we assign
any order (\textit{ANY}) traversal strategy (path $P_{11}$).
This is the case for traversal \lstinline|initT| in the post-dominator analysis
(\autoref{lst:dominators}).

For traversals that are data-flow sensitive, such as \lstinline|domT|
in the post-dominator analysis,
we further check the \graphprop{} property of the input graphs.
The loop sensitivity property is only applicable to graphs with loops, thus, is
considered last.

\textbf{Sequential Graphs (paths $P_1$ and $P_2$)}. In case of sequential
graphs, each node has at most one predecessor or one successor, and the data
flows to any node are from either its predecessor or its successor, not both.
Thus, the most efficient strategy should traverse the graph in the 
direction of the data flow, so that the traversal can finish
with only one iteration, and no fixpoint nor worklist is needed.
We use traversal strategy \textit{INC} (leaf node $P_1$) for 
\lstinline|FORWARD| analysis and \textit{DEC} (leaf node $P_2$) for 
\lstinline|BACKWARD| analysis. 
	
	
\textbf{Graphs with branches (paths $P_3$ and $P_4$)}.
These graphs contain branches but no loops, hence a node may have more than one
successor or predecessor. If the traversal under consideration is data-flow
sensitive, depending on the traversal direction, it requires the output of
either the successors or predecessors to compute the output for any node.
For \lstinline|FORWARD| traversal direction, we need to visit all predecessors
of any node prior to visiting the node and this traversal order is given by the
post-order (\textit{RPO}) and hence we select \textit{RPO}. For
\lstinline|BACKWARD| traversal direction, we need to visit all successors of any
node prior to visiting the node and this traversal order is given by the reverse
post-order (\textit{PO}), hence we select \textit{PO}.

\textbf{Graphs with loops (paths $P_5$ to $P_{10}$)}. These graphs contain both
branches and loops, hence we need to check if the traversal is sensitive to the
loop (the loop sensitive property). The decision for \lstinline|domT| falls into
these paths because it has loops.

\textbf{Loop sensitive (paths $P_9$ and $P_{10}$)}. When the traversal is
loop sensitive, for correctly propagating the output, the traversal visits
nodes multiple times until a fix point condition is satisfied (user may
provide a fix point function). 
We adopt a worklist-based traversal strategy because it visits only required 
nodes by maintaining a list of nodes left to be processed. 
For picking the best order of traversing nodes in a graph to initialize the worklist, 
we further investigate
the traversal direction. We know that, for \lstinline|FORWARD| traversals, 
\textit{RPO} strategy gives the best order of nodes and for
\lstinline|BACKWARD| traversals, \textit{PO} strategy gives
the best order of nodes, we pick worklist strategy with reverse post-order
\textit{WRPO} for \lstinline|FORWARD| and worklist strategy with post-order
\textit{WPO} for \lstinline|BACKWARD| traversal directions.

\textbf{Loop insensitive (paths $P_5$ to $P_8$)}. When the traversal is
loop insensitive, the selection problem reduces to the sequential and branch case
that is discussed previously, because for loop insensitive traversal, the
loops in the input graph are irrelevant and what remains relevant
is the existence of the branches. 
For example, \lstinline|domT| is loop insensitive, thus, is a case. 
As the graph contains branches, the next property to be checked is the traversal
direction. Since the traversal direction for \lstinline|domT| is \lstinline|BACKWARD|,
we pick \textit{PO} traversal strategy for \lstinline|domT|, as shown
by the path $P_6$ in \fignref{fig:decision-diagram}. 


\subsection{Optimizing the Selected Traversal Strategy}
\label{subsec:optimizations}

The properties of the analysis and the input graph not only help us select the
best traversal strategy but also help perform several optimizations to the
selected traversal strategies. 

Running an analysis on a cyclic graph may require multiple iterations 
to compute the fixpoint solution. At least one last traversal 
is required to check that the results of the last
two traversals have not changed. This
additional traversal and the fixpoint checks at nodes can be eliminated based on
the selected traversal strategy.

For data-flow insensitive traversals, the traversal outputs of
nodes do not depend on the outputs of other nodes, hence both the additional
traversal and the fixpoint checks can be eliminated irrespective of the graph
cyclicity (decision path $P_{11}$ in 
\fignref{fig:decision-diagram}).

In case of data-flow sensitive traversals, the traversal output of a node is
computed using the traversal outputs of other nodes (predecessors or
successors).
For data-flow sensitive analysis on an acyclic graph, the additional traversal and
fixpoint checks are not needed if selected traversal strategy guarantees that
the nodes whose outputs are required to compute the output of a node are visited
prior to visiting the node (paths $P_1$--$P_4$).
In case of data-flow, loop sensitive traversals on graphs with loops, the
optimization does not apply as no traversal strategy can guarantee that the
nodes whose outputs are required to compute the output of a node are visited
prior to visiting the node. Due to the presence of loops, \shortname
selects the worklist-based traversal strategy (paths $P_9$ and $P_{10}$).
However, in case of data-flow and loop in-sensitive traversals on graphs with
loops, the loops can be ignored since the traversal is insensitive to it and the
additional traversal and fixpoint checks can be eliminated if the selected
traversal strategy guarantees that the nodes, whose outputs are required to
compute the output of a node, are visited prior to visiting the node (paths
$P_5$--$P_8$).

\subsection{Extracting Analysis Properties}
\label{sec:language}
In \secnref{sec:compute-properties} we described a set of static properties of
the analysis that influence the traversal strategy selection, we now provide a
technique to automatically extract them by analyzing the analysis source code.
Before that, we first describe the DSL that we chose (Boa) and the
extensions that we added to support effective extraction of the static
properties of the analysis programs written in this DSL.\footnote{Our reason
for selecting the Boa DSL is that it already provides an infrastructure to
facilitate ultra-large scale analysis (several hundred millions of methods).}

The Boa DSL provides features to access source code files, revisions, classes,
methods, statements, expressions, etc~\cite{boa-gpce}. It also provides CFGs of methods over
which the analysis can be performed. We extend Boa DSL to support writing
iterative data-flow analysis.\footnote{The DSL extensions that we added
are nothing specific to Boa, instead are common constructs that can be found in any
framework that implements iterative data-flow style analysis, for example, Soot
analysis framework.}
A typical iterative data-flow analysis is described using: i) a traversal
function (containing instructions to compute the analysis output for any CFG
node), ii) a fixpoint function (that defines how to merge the outputs of
predecessors or successors, if the analysis requires), and iii) the traversal
direction.\footnote{The common data structures to store the analysis outputs at
CFG nodes is already available in Boa. The Boa DSL and our extension was
sufficient to express all the analyses in our evaluation dataset.} We now
describe our extension using a sample analysis shown in Listing
\ref{lst:dominators}.

%
%
%
%

\begin{lstlisting}[basicstyle=\footnotesize\ttfamily, numbers=left, numbersep=-8pt, 
escapechar=@, caption={Post-dominator analysis: an example source code analysis
expressed using our DSL extension.}, label={lst:dominators}] 
		allNodes: Set<int>;
		initT := traversal(n: Node) {
			add(allNodes, n.id);
		}
		domT := traversal(n: Node): Set<int> { 
			Set<int> dom;
			if (output(n, domT) != null) dom = output(n, domT);
			else if (node.id == exitNodeId)	dom = {};
			else	dom = allNodes;
			foreach (s : n.succs) 
				dom = intersection(dom, output(s, domT)) 
			add(dom, n.id); 
			return dom; 
		} 
		fp := fixp(Set<int> curr, Set<int> prev): bool {
			if (equals(curr, prev))	return true;
			return false;
		}
		traverse(g, initT, ITERATIVE); @\label{line:traverse}@
		traverse(g, domT, BACKWARD, fp); 				
\end{lstlisting}

\para{Post-dominator analysis} 
The post-dominator analysis is a backward control flow analysis that collects
node ids of all nodes that post-dominates every node in the CFG~\cite{compilers}.
This analysis can be expressed using our DSL extension as shown in Listing
\ref{lst:dominators} that defines two traversals \lstinline|initT|
(lines 2-4) and \lstinline|domT| (lines 5-14). A traversal is defined using a
special \lstinline|traversal| block:
\begin{lstlisting} 
t := traversal(n : Node) : T { tbody }
\end{lstlisting}
In this traversal block definition, \lstinline|t| is the name of the traversal
that takes a single parameter \lstinline|n| representing the graph node that is
being visited. A traversal may define a return type \lstinline|T| representing
the output type. The output type can be a primitive or a collection data type.
In our example, the traversal \lstinline|initT| does not define any output for
CFG nodes, whereas, the traversal \lstinline|domT| defines an output of type
\lstinline|Set<int>| for every node in the CFG. A block of code that generates
the traversal output at a graph node is given by \lstinline|tbody|. The
\lstinline|tbody| may contain common statements and expressions, such as
variable declarations, assignments, conditional statements, loop statements, and
method calls, along with some special expressions discussed in this section.
These traversals are invoked by using a special \lstinline|traverse| expression
(lines 19 and 20).
\begin{lstlisting}
traverse(g, t, d, fp)
\end{lstlisting}

A \lstinline|traverse| expression takes four parameters: \lstinline|g| is the
graph to be traversed, \lstinline|t| is the traversal to be invoked,
\lstinline|d| is the traversal direction and \lstinline|fp| is an optional 
variable name of the user-defined fixpoint function. A
traversal direction is a value from the set \{\lstinline|FORWARD|,
\lstinline|BACKWARD|, \lstinline|ITERATIVE|\}.
\lstinline|ITERATIVE| represents a sequential analysis that visits nodes as they appear 
in the nodes collection.

Lines 15-18 defines a
fixpoint function using \lstinline|fixp| block, used in the
\lstinline|traverse| expression in line 20.
\lstinline|fixp| is a keyword for defining a
fixpoint function. A fixpoint function can take 0 or more parameters, and 
must always return a boolean. A fixpoint function can be assigned a name, which
can be passed in the \lstinline|traverse| expression.

Line 1 defines a variable \lstinline|allNodes| of type \lstinline|Set<int>|
which defines a collection type \lstinline|Set| with elements of type
\lstinline|int|. The Boa DSL provides primitive and collection data types.
Collection types include: \lstinline|Set and Seq|, where \lstinline|Set| is a
collection with unique and unordered elements, whereas, \lstinline|Seq| is a
collection with non unique but ordered elements. Important set of operations
that can be performed on collection types are \texttt{add(C1, e)} that adds an
element e to collection C1, \texttt{addAll(C1, C2)} that adds all elements from
collection C2 to collection C1, \texttt{remove(C1, e)} that removes an element e
from Collection C1, \texttt{removeAll(C1, C2)} that removes all elements present
in collection C2 from collection C1, \texttt{union(C1, C2)} that returns union
of the elements in C1 and C2 and \texttt{intersection(C1, C2)} that returns
intersection of the elements in C1 and C2. Line 3 in our example uses an
operation \lstinline|add| on collection \lstinline|allNodes|.

A usage of special expression
\lstinline|output(n, domT)| can be seen in~line 7. This is used for querying the
traversal output associated with a graph node \lstinline|n|, in the traversal
\lstinline|domT|. The
traversal \lstinline|domT| defines an output of type \lstinline|Set<int>| for
every node in the CFG. For managing the analysis output of nodes,
\lstinline|domT| traversal maintains an internal map that contains analysis
output for every node, which can be queried using \lstinline|output(n, domT)|.
A pre-defined variable \lstinline|g| that represents the CFG is used in the
\lstinline|traverse| expressions in lines 19--20.

\begin{figure}[ht!]
\centering
\includegraphics[width=\linewidth]{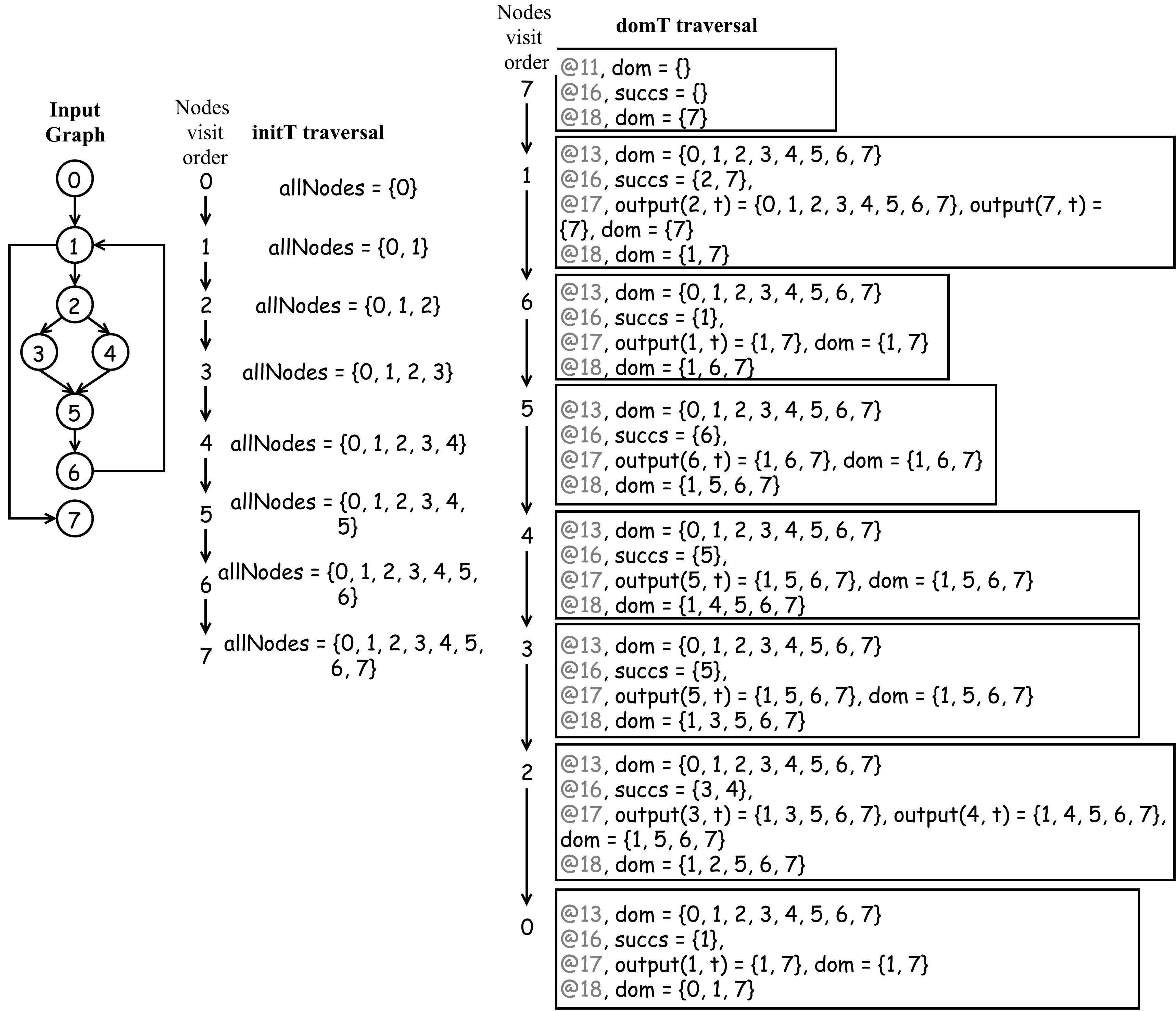}
\caption{Running example of applying the post-dominator analysis on an input
graph containing branch and loop.}
\label{fig:running-example}
\end{figure}

\fignref{fig:running-example} takes an example graph, and shows the results of
\lstinline|initT| and \lstinline|domT| traversals. Our example graph is a CFG
containing seven nodes with a branch and a loop. The \lstinline|initT| traversal
visits nodes sequentially and adds node \lstinline|id| to the collection
\lstinline|allNodes|. The \lstinline|domT| traversal visits nodes in the
post-order\footnote{The traversal strategies chosen for \lstinline|initT| and
\lstinline|domT| traversals are explained in \secnref{sec:decision-tree}.} and
computes a set of nodes that post-dominate every visited node (as indicated by
the set of node ids). For instance, node 7 is post-dominated by itself, hence
the output at node 7 is \{7\}.
In \fignref{fig:running-example}, under \lstinline|domT| traversal, for each
node visited, we show the key intermediate steps indicated by $@$ line number.
These line numbers correspond to the line numbers shown in Listing
\ref{lst:dominators}. We will explain the intermediate results while visiting
node 2. In the \lstinline|domT| traversal, at line 9, the output set
\lstinline|dom| is initialized using \lstinline|allNodes|, hence \lstinline|dom|
= \{0, 1, 2, 3, 4, 5, 6, 7\}. At line 10, node 2 has two successors: \{3, 4\}.
At line 11, \lstinline|dom| is updated by performing an
\lstinline|intersection| operation using the outputs of successors 3 and 4. Since the
outputs of 3 and 4 are \{1, 3, 5, 6, 7\} and \{1, 4, 5, 6, 7\}, respectively, 
\lstinline|dom| set becomes
\{1, 5, 6, 7\}. At line 12, the \lstinline|id| of the visited node is added to the
\lstinline|dom| set and it becomes \{1, 2, 5, 6, 7\}.
Hence, the post-dominator set for node 2 is \{1, 2, 5, 6, 7\}. 
The post-dominator sets for other nodes are calculated similarly.

In the next two subsections, we describe how data-flow and loop sensitivity
properties can be extracted from the analysis expressed using our extension. The
traversal direction property need not be extracted as it is directly provided by
the analysis writer.

\subsubsection{Computing Data-Flow Sensitivity}
\label{sec:df-algo}

To determine the data-flow sensitivity property of a traversal, the operations
performed in the traversal needs to be analyzed to check if the output of a node
is computed using the outputs of other nodes. In our DSL extension, the only way
to access the output of a node in a traversal is via \lstinline|output()|
expression, hence given a traversal \lstinline|t| := \lstinline|traversal(n : Node)| :
\lstinline|T| \{ \lstinline|tbody| \} as input, \algoref{algo:flow-sensitive}
parses the statements in the traversal body \lstinline|tbody| to identify method
calls of the form \lstinline|output(n', t')| that fetches the output of a node
\lstinline|n'| in the traversal \lstinline|t'|. If such method calls exist,
they are further investigated to determine if \lstinline|n'| does not point to
\lstinline|n| and \lstinline|t'| points to \lstinline|t|.
If this also holds, it means that the traversal output for the current node
\lstinline|n| is computed using the traversal outputs of other nodes
(\lstinline|n'|) and hence the traversal is data-flow sensitive.
For performing the points-to check, \algoref{algo:flow-sensitive} assumes that
an alias environment is computed by using \textit{must alias}
analysis~\cite{must-alias} which computes all names in the \lstinline|tbody|
that must alias each other at any program point. The must alias information
ensures that \algoref{algo:flow-sensitive} never classifies a data-flow
sensitive traversal as data-flow insensitive.
The control and loop statements
in the \lstinline|tbody| do not have any impact on \algoref{algo:flow-sensitive}
for computing the data-flow sensitivity property.

\begin{algorithm}[ht!]
\caption{Algorithm to detect data-flow sensitivity}
\label{algo:flow-sensitive}
\KwIn{\lstinline|t := traversal(n : Node) : T \{ tbody \}|}
\KwOut{\textit{true/false}}
$A$ $\leftarrow$ \textit{getAliases}(\lstinline|tbody|, \lstinline|n|)\;
\ForEach{ \lstinline|stmt| $\in$ \lstinline|tbody|}{
	\If{$stmt$ $=$ \lstinline|output(n', t')|}{
		\If{\lstinline|t'| $==$ \lstinline|t| and \lstinline|n'| $\notin$ $A$ }{
			return \textit{true}\;
		}
	}
	
}
return \textit{false}\;
\end{algorithm}

\subsubsection{Computing Loop Sensitivity}
\label{sec:ls-algo}
In general, computing the loop sensitivity property statically is challenging in
the absence of an input graph. However, the constructs and operations provided in
our DSL extension enable the static inference of this property.

A traversal is loop sensitive if the output of a node in any two successive 
iterations either shrinks or expands. 
We analyze the operations performed in the traversal to determine if the 
traversal output expands or shrinks in the subsequent iterations. 
The operations \lstinline|add|, \lstinline|addAll|,
and \lstinline|union| always expand the output while 
\lstinline|remove|, \lstinline|removeAll|, and \lstinline|intersection| always
shrink the output.

\begin{algorithm}
\caption{Algorithm to detect loop sensitivity}
\label{algo:loop-sensitive}
\vspace{-10pt}
\begin{multicols}{2} 
\KwIn{t := \lstinline|traversal(n: Node) : T \{ tbody \}|}
\KwOut{\textit{true/false}}
$V$ $\leftarrow$ \{\} // a set of output variables related to n\;
$V'$ $\leftarrow$ \{\} // a set of output variables not related to n\; 
\lstinline|expand| $\leftarrow$ \textit{false}\; 
\lstinline|shrink| $\leftarrow$ \textit{false}\;
\lstinline|gen| $\leftarrow$ \textit{false}\; 
\lstinline|kill| $\leftarrow$ \textit{false}\;
$A$ $\leftarrow$ \textit{getAliases($n$)}\; 
\ForEach{ \lstinline|s| $\in$ \lstinline|tbody| }{
	\If{ \lstinline|s| is \lstinline|v| = \lstinline|output|($n'$,
	$t'$)}{ \If{$t'$ $==$ \lstinline|t|}{
			\eIf{$n'$ $\in$ $A$}{
				$V$ $\leftarrow$ $V$ $\cup$ \textit{\lstinline|v|}\;
			}{
				$V'$ $\leftarrow$ $V'$ $\cup$ \textit{\lstinline|v|}\;					
			}
		}
	}
}
\ForEach{ \lstinline|s| $\in$ \lstinline|tbody| }{
		\If{ \lstinline|s| $=$ \lstinline|union|($c_1$, $c_2$) }{
			\lIf{ ($c_1$ $\in$ $V$ and $c_2$ $\in$ $V'$) $||$ ($c_1$ $\in$ $V'$ and $c_2$
			$\in$ $V$)}{ 
				\lstinline|expand| $\leftarrow$ \textit{true}
			}
		}
		\If{ \lstinline|s| $=$ \lstinline|intersection|($c_1$, $c_2$)}{
			\lIf{($c_1$ $\in$ $V$ and $c_2$ $\in$ $V'$) $||$ ($c_1$ $\in$ $V'$
			and $c_2$ $\in$ $V$)}{ 
				\lstinline|shrink| $\leftarrow$ \textit{true}
			}
		}
		\If{ \lstinline|s| $=$ \lstinline|add|($c_1$, $e$) $||$	
					\lstinline|s| $=$ \lstinline|addAll|($c_1$, $c_2$)}{ 
			\If{$c_1$ $\in$ $V$}{ 
				\lstinline|gen| $\leftarrow$ \textit{true}\;
			}
		}
		\If{ \lstinline|s| $=$ \lstinline|remove|($c_1$, $e$) $||$
					\lstinline|s| $=$ \lstinline|removeAll|($c_1$, $c_2$)}{ 
			\If{$c_1$ $\in$ $V$}{ 
				\lstinline|kill| $\leftarrow$ \textit{true}\;
			}
		}
}
\leIf{ (\lstinline|expand| and \lstinline|gen|) $||$ (\lstinline|shrink| and
\lstinline|kill|) }{ return \textit{true}\;
}{
	return \textit{false}
}
\end{multicols}
\end{algorithm}
%

Given a traversal \lstinline|t| := \lstinline|traversal(n : Node)| :
\lstinline|T| \{ \lstinline|tbody| \}, \algoref{algo:loop-sensitive} determines
the loop sensitivity of \lstinline|t|.
\algoref{algo:loop-sensitive} investigates the statements in the
\lstinline|tbody| to determine if the traversal outputs of nodes in multiple
iterations either expand or shrink. For doing that, first it parses the
statements to collect all output variables related and not related to input node
\lstinline|n| using the must alias information as in
\algoref{algo:flow-sensitive}. This is determined in lines 8-14, where all
output variables are collected (output variables are variables that gets
assigned by the \lstinline|output| operation) and added to two sets $V$ (a set
of output variables related to n) and $V'$ (a set of output variables not
related to n).
Upon collecting all output variables, \algoref{algo:loop-sensitive} makes
another pass over all statements in the \lstinline|tbody| to identify six kinds
of operations: \lstinline|union|, \lstinline|intersection|, \lstinline|add|,
\lstinline|addAll|, \lstinline|remove|, and \lstinline|removeAll|. 
In lines 16--17, the algorithm looks for \lstinline|union| operation, where one
of the variables involved is an output variables related to \lstinline|n| and
the other variable involved is not related to \lstinline|n|. These conditions
are simply the true conditions for the data-flow sensitivity, where the output
of the current node is computed using the outputs of other nodes (neighbors).
Similarly, in lines 18--19, the algorithm looks for \lstinline|intersection|
operation. Lines 20--25 identifies add and remove operations that add or
remove elements from the output related to node \lstinline|n|. Finally, if
there exist \lstinline|union| and \lstinline|add| operations, the output of a
node always expands, and if there exist \lstinline|intersection| and
\lstinline|remove| operations, the output of a node always shrinks. For a
data-flow traversal to be loop sensitive, the output of nodes must either expand
or shrink, not both (lines 26--27).

\section{Empirical Evaluation}
\label{sec:eval}

We conducted an empirical evaluation on a set of 21 basic source code analyses
on two public massive code datasets to evaluate several factors of \shortname.
First, we show the benefit of using \shortname over standard strategies
by evaluating the \emph{reduction in running times} of \shortname
over the standards ones (\secref{sec:performance-gain}).
Then, we evaluate the correctness of the analysis results using \shortname 
to show that the decision analyses and optimizations in \shortname
do not affect the correctness of the source code analyses
(\secref{sec:soundness}).
We also evaluate the precision of our selection algorithm by measuring how often
\shortname selects the most time-efficient traversal
(\secref{sec:prediction-accuracy}). 
We evaluate how the different components of \shortname and different
kinds of static and runtime properties impact the overall performance 
in \secref{sec:analysis-traversal-opt}.
Finally, we show practical uses of \shortname in 
three applications in \secref{sec:case-studies}.

%
%
%

\subsection{Analyses, Datasets and Experiment Setting}
\label{sec:exp-settings}

\begin{table}
  \centering
  \caption{List of source code analyses and properties of their involved 
	traversals. \footnotesize Ts: total number of traversals. $t_i$: properties of
	the $i$-th traversal. \emph{Flw}: data-flow sensitive. \emph{Lp}: loop
	sensitive.
	\emph{Dir}: traversal direction where ---, $\rightarrow$ and $\leftarrow$ 
	mean iterative, forward and backward, respectively. \cmark and \xmark~ for
	\emph{Flw} and \emph{Lp} indicates whether the property is \textit{true} or
	\textit{false}.}
	\scriptsize
	\setlength{\tabcolsep}{2.4pt}
\begin{tabular}{rrrcccccrccc}
\toprule
      & \multicolumn{1}{c}{Analysis} & \multicolumn{1}{c}{Ts} & \multicolumn{3}{c}{$t_1$} & \multicolumn{3}{c}{$t_2$} & \multicolumn{3}{c}{$t_3$} \\
\cmidrule{4-6} \cmidrule(lr){7-9} \cmidrule(lr){10-12}
      & \multicolumn{1}{c}{} & \multicolumn{1}{c}{} & Flw   & Lp  & Dir & Flw   & Lp  & \multicolumn{1}{c}{Dir} & Flw   & Lp  & Dir \\
\midrule
1     & Copy propagation (CP) & 3     & \xmark & \xmark & ---   & \cmark & \cmark & $\rightarrow$ & \xmark & \xmark & --- \\
2     & Common sub-expression detection (CSD) & 3     & \xmark & \xmark & ---   & \cmark & \cmark & $\rightarrow$ & \xmark & \xmark & --- \\
3     & Dead code (DC) & 3     & \xmark & \xmark & ---   & \cmark & \cmark & $\leftarrow$ & \xmark & \xmark & --- \\
4     & Loop invariant code (LIC) & 3     & \xmark & \xmark & ---   & \cmark & \cmark & $\rightarrow$ & \xmark & \xmark & --- \\
5    & Upsafety analysis (USA) & 3     & \xmark & \xmark & ---   & \cmark & \cmark & $\rightarrow$ & \xmark & \xmark & --- \\
6    & Valid FileReader (VFR) & 3     & \xmark & \xmark & ---   & \cmark & \cmark & $\rightarrow$ & \xmark & \xmark & --- \\
7    & Mismatched wait/notify (MWN) & 3     & \xmark & \xmark & ---   & \cmark & \cmark & $\rightarrow$ & \xmark & \xmark & --- \\
8     & Available expression (AE)  & 2     & \xmark & \xmark & ---   & \cmark & \cmark & $\rightarrow$ & \cellcolor{lightgray} & \cellcolor{lightgray} & \cellcolor{lightgray} \\
9     & Dominator (DOM)  & 2     & \xmark & \xmark & ---   & \cmark & \xmark & $\rightarrow$ & \cellcolor{lightgray} & \cellcolor{lightgray} & \cellcolor{lightgray} \\
10     & Local may alias (LMA) & 2     & \xmark & \xmark & ---   & \cmark & \cmark & $\rightarrow$ & \cellcolor{lightgray} & \cellcolor{lightgray} & \cellcolor{lightgray} \\
11     & Local must not alias (LMNA) & 2     & \xmark & \xmark & ---   & \cmark & \cmark & $\rightarrow$ & \cellcolor{lightgray} & \cellcolor{lightgray} & \cellcolor{lightgray} \\
12     & Live variable (LV) & 2     & \xmark & \xmark & ---   & \cmark & \cmark & $\leftarrow$ & \cellcolor{lightgray} & \cellcolor{lightgray} & \cellcolor{lightgray} \\
13    & Nullness analysis (NA) & 2     & \xmark & \xmark & ---   & \cmark & \cmark & $\rightarrow$ & \cellcolor{lightgray} & \cellcolor{lightgray} & \cellcolor{lightgray} \\
14    & Post-dominator (PDOM) & 2     & \xmark & \xmark & ---   & \cmark & \xmark & $\leftarrow$ & \cellcolor{lightgray} & \cellcolor{lightgray} & \cellcolor{lightgray} \\
15    & Reaching definition (RD)  & 2     & \xmark & \xmark & ---   & \cmark & \cmark & $\rightarrow$ & \cellcolor{lightgray} & \cellcolor{lightgray} & \cellcolor{lightgray} \\
16    & Resource status (RS) & 2     & \xmark & \xmark & ---   & \cmark & \cmark & $\rightarrow$ & \cellcolor{lightgray} & \cellcolor{lightgray} & \cellcolor{lightgray} \\
17    & Very busy expression (VBE)  & 2     & \xmark & \xmark & ---   & \cmark & \cmark & $\leftarrow$ & \cellcolor{lightgray} & \cellcolor{lightgray} & \cellcolor{lightgray} \\
18    & Safe Synchronization (SS) & 2     & \xmark & \xmark & ---   & \cmark & \cmark & $\rightarrow$ & \cellcolor{lightgray} & \cellcolor{lightgray} & \cellcolor{lightgray} \\
19    & Used and defined variable (UDV) & 1     & \xmark & \xmark & ---   & \cellcolor{lightgray} & \cellcolor{lightgray} & \multicolumn{1}{c}{\cellcolor{lightgray}} & \cellcolor{lightgray} & \cellcolor{lightgray} & \cellcolor{lightgray} \\
20    & Useless increment in return (UIR) & 1     & \xmark & \xmark & ---   & \cellcolor{lightgray} & \cellcolor{lightgray} & \multicolumn{1}{c}{\cellcolor{lightgray}} & \cellcolor{lightgray} & \cellcolor{lightgray} & \cellcolor{lightgray} \\
21    & Wait not in loop (WNIL) & 1     & \xmark & \xmark & ---   & \cellcolor{lightgray} & \cellcolor{lightgray} & \multicolumn{1}{c}{\cellcolor{lightgray}} & \cellcolor{lightgray} & \cellcolor{lightgray} & \cellcolor{lightgray} \\
\bottomrule
\end{tabular}%
  \label{tab:analysis-table}%
\end{table}%

\subsubsection{Analyses.} We collected source code analyses that traverse CFGs
from textbooks and tools. We also ensured that the 
analyses list covers all the static properties discussed in  
\secnref{sec:compute-properties}, i.e., data-flow sensitivity, loop sensitivity and 
traversal direction (forward, backward and iterative).
We ended up with 21 source code analyses as shown in 
\tabref{tab:analysis-table}. They include 10 basic ones (analyses 1, 2, 8, 9, 
10, 11, 12, 14, 15 and 19) from textbooks~\cite{compilers, programanalysis} and 11 
others for detecting source code bugs, and code smells from the Soot 
framework~\cite{soot-cascon1999} (analyses 3, 4, 5, 13, 17 and 18), and 
FindBugs tool~\cite{findbugs-paste2007} (analyses 6, 7, 16, 20 and 21). 
\tabref{tab:analysis-table} also shows the number of traversals each analysis
contains and their static properties as described in
\secref{sec:compute-properties}.
All analyses are intra-procedural.
We implemented all twenty one of these analysis in Boa using the constructs
described in \secref{sec:language}.\footnote{Our implementation infrastructure
Boa currently supports only method-level analysis, however our technique should
be applicable to inter-procedural CFGs.}

\begin{table}
\scriptsize
\centering
\caption{Statistics of the generated control flow graphs.}
\setlength{\tabcolsep}{2pt}
\begin{tabular}{lrrrrrr}
\toprule
\multicolumn{1}{c}{Dataset} & \multicolumn{1}{c}{All graphs} & \multicolumn{1}{c}{Sequential} & \multicolumn{1}{c}{Branches} & \multicolumn{3}{c}{Loops} \\
\cmidrule(lr){5-7}      &       &       &       & \multicolumn{1}{c}{All graphs} & \multicolumn{1}{c}{Branches} & \multicolumn{1}{c}{No branches} \\
\midrule
DaCapo & 287K & 186K (65\%) & 73K (25\%) & 28K (10\%) & 21K (7\%) & 7K (2\%)\\
GitHub & 161,523K & 111,583K (69\%) & 33,324K (21\%) & 16,617K (10\%) & 11,674K (7\%) & 4,943K (3\%) \\
\bottomrule
\end{tabular}%
\label{tab:dataset-table}%
\end{table}%

\begin{table}
	\centering
	\small
	\caption{Time contribution of each phase (in miliseconds).}
\setlength{\tabcolsep}{4.5pt}
\begin{tabular}{rrrrrrrr}
\toprule
Analysis & \multicolumn{2}{c}{Avg. Time} & Static & \multicolumn{4}{c}{Runtime} \\
\cmidrule(lr){2-3} \cmidrule(lr){5-8}      & DaCapo & GitHub &       & \multicolumn{2}{c}{DaCapo} & \multicolumn{2}{c}{GitHub} \\
\cmidrule(lr){5-6} \cmidrule(lr){7-8}      &       &       &       & Avg.  & Total & Avg.  & Total \\
\midrule
\multicolumn{1}{l}{CP} & 0.21 &    0.008   &       53  &   0.21    &  62,469     &    0.008   & 1359K \\
\multicolumn{1}{l}{CSD} & 0.19 &  0.012     &       60  &     0.19  &   56,840    &   0.012    &  1991K \\
\multicolumn{1}{l}{DC} & 0.19 &   0.010    &       45  &    0.19   &  54,822     &   0.010    &  1663K \\
\multicolumn{1}{l}{LIC} & 0.21 &  0.006     &       69  &    0.20   &    60,223   &    0.006   &  992K \\
\multicolumn{1}{l}{USA} & 0.19 &   0.006    &       90  &   0.19    &   54,268    &  0.009     &  1444K \\
\multicolumn{1}{l}{VFR} & 0.18 &    0.007   &       42  &   0.18    &     52,483  &   0.007    &  1142K \\
\multicolumn{1}{l}{MWN} & 0.18 &    0.006   &       36  &   0.18    &     52,165  &   0.006    &  1109K \\
\multicolumn{1}{l}{AE} & 0.18 &    0.007   &       43  &   0.18    &     53,290  &   0.007    &  1169K \\
\multicolumn{1}{l}{DOM} & 0.21 &    0.008   &       35  &    0.21   &    62,416    &  0.008     &  1307K \\
\multicolumn{1}{l}{LMA} & 0.18 &    0.008   &       76  &   0.18    &     52,483  &   0.008    &  1346K \\
\multicolumn{1}{l}{LMNA} & 0.18 &    0.008   &       80  &     0.18  &   53,182    &  0.008     &  1407K\\
\multicolumn{1}{l}{LV} & 0.17 &   0.007    &       32  &    0.17   &     49,231  &   0.007    &  1273K\\
\multicolumn{1}{l}{NA} & 0.16 &   0.008    &       64  & 0.16      &    46,589   &   0.008    &  1398K\\
\multicolumn{1}{l}{PDOM} & 0.20 &    0.012   &       34  &   0.20    &   57,203    &   0.012    &  2040K\\
\multicolumn{1}{l}{RD} & 0.20 &   0.007    &       48  &    0.20   &   57,359    &  0.007     &  1155K\\
\multicolumn{1}{l}{RS} & 0.16 &   0.006    &       28  &   0.16    &   46,367    &   0.006    &  996K\\
\multicolumn{1}{l}{VBE} & 0.17 &  0.006     &       44  &   0.17    &   49,138    &   0.006    &  1062K\\
\multicolumn{1}{l}{SS} & 0.17 &  0.006     &       32  &   0.17    &   48,990    &   0.006    &  1009K\\
\multicolumn{1}{l}{UDV} & 0.14 &  0.005     &       10  &    0.14   &  41,617     &  0.005     &  928K\\
\multicolumn{1}{l}{UIR} & 0.14 &   0.006    &       14  &   0.14    &   41,146    &   0.006    &  1020K\\
\multicolumn{1}{l}{WNIL} & 0.14 &   0.007    &       15  &   0.14    &  41,808     &  0.007     &  1210K\\
\bottomrule
\end{tabular}%

	\label{tab:time}%
\end{table}%

\subsubsection{Datasets.} We ran the analyses on two datasets: DaCapo 9.12
benchmark~\cite{blackburn2006dacapo}, and a large-scale dataset containing
projects from GitHub.
DaCapo dataset contains the source code of 10 open source Java projects:
Apache Batik, Apache FOP, Apache Aurora, Apache Tomcat, Jython, Xalan-Java, PMD,
H2 database, Sunflow and Daytrader. GitHub dataset contains the source code of
more than 380K Java projects collected from GitHub.com. Each method in the
datasets was used to generate a control flow graph (CFG) on which the analyses
would be run. The statistics of the two datasets are shown in
\tabref{tab:dataset-table}. 
Both have similar distributions of CFGs over
graph \graphprop{} (i.e., sequential, branch, and loop). 

\subsubsection{Setting.}\label{subsubsec:setting} We compared \shortname 
against six standard traversal strategies in
\secref{sec:candidates}: DFS, PO, RPO, WPO, WRPO and ANY. The running time for
each analysis is measured from the start to the end of the analysis. The running
time for \shortname also includes the time for computing the static and
runtime properties, making the traversal strategy decision, optimizing it and
then using the optimized traversal strategy to traverse the CFG, and run the
analysis.
The analyses on DaCapo dataset were run on  a single machine with 24 GB of RAM
and 24 cores running Linux 3.5.6-1.fc17 kernel. Running analyses on GitHub
dataset on a single machine would take weeks to finish, so we ran them on a
cluster that runs a  standard  Hadoop 1.2.1 with 1 name and job
tracker node, 10 compute nodes with totally 148 cores, and 1 GB of RAM for each
map/reduce task.


\subsection{Running Time and Time Reduction}
\label{sec:performance-gain}

We first report the running times and then study the reductions (or speedup)
against standard traversal strategies.

\subsubsection{Running Time}

\tabref{tab:time} shows the running times for 21 analyses on the two datasets.
On average (column \textbf{Avg. Time}), each analysis took 0.14--0.21 ms and
0.005--0.012 ms to analyze a graph in Dacapo and GitHub datasets, respectively.
The variation in the average analysis time is mainly due to the difference in
the machines used to run the analysis for DaCapo and GitHub datasets.
Also, the graphs in DaCapo are on
average much larger compared to GitHub.
%
Columns \textbf{Static} and \textbf{Runtime} show the time contributions for 
different components of \shortname: the time for determining the 
static properties of each analysis which is done once for each analysis, and 
the time for constructing the CFG of each method and traversing the CFG which 
is done once for every constructed CFG. 
We can see that the time for collecting static information is negligible, 
less than 0.2\% for DaCapo dataset and less than 0.01\% for GitHub 
dataset, when compared to the total runtime information collection time, as it is performed only 
once per traversal. When compared to the average runtime information collection time, the static 
time is quite significant. However, the overhead introduced by static 
information collection phase diminishes as the number of CFGs increases and 
becomes insignificant when running on those two large datasets. This result 
shows the benefit of \shortname when applying on large-scale
analysis.

\subsubsection{Time Reduction}

\begin{figure*}%
\centering
\small
\subfloat[Time reduction for each analysis.\label{tab:reduction-analyses}] {
\begin{tabular}{lrrrrrr|rrrrrr}
\toprule
\multicolumn{1}{l}{Analysis} & \multicolumn{6}{c|}{DaCapo}                    & \multicolumn{6}{c}{GitHub} \\
\cmidrule(lr){2-7} \cmidrule(lr){8-13}
      & DFS   & PO    & RPO   & WPO   & WRPO  & ANY    & DFS   & PO    & RPO   & WPO   & WRPO  & ANY \\
\midrule
\multicolumn{1}{l}{CP} & \cellcolor{lightblack}17\%  & \cellcolor{lightgreen}83\%  &  \cellcolor{lightblue}9\%   & \cellcolor{lightgreen}66\%  &  \cellcolor{lightblack}11\%   & \cellcolor{lightgreen}72\%  & \cellcolor{lightblack}17\%  & \cellcolor{lightgreen}88\%  &  \cellcolor{lightblack}12\%   & \cellcolor{lightgreen}80\%  &  \cellcolor{lightblue}5\%   & \cellcolor{lightgreen}82\% \\
\multicolumn{1}{l}{CSD} & \cellcolor{lightblack}41\%  & \cellcolor{lightgreen}93\%  & \cellcolor{lightblack}39\%  & \cellcolor{lightgreen}74\%  &  \cellcolor{lightblue}4\%   & \cellcolor{lightgreen}89\%  & \cellcolor{lightblack}31\%  & --  & \cellcolor{lightblack}24\%  & --  &  \cellcolor{lightblack}12\%   & -- \\
\multicolumn{1}{l}{DC} & \cellcolor{lightblack}41\%  & \cellcolor{lightblack}30\%  & \cellcolor{lightgreen}89\%  &  \cellcolor{lightblue}7\%   & \cellcolor{lightgreen}64\%  & \cellcolor{lightgreen}81\%  & \cellcolor{lightblack}25\%  & \cellcolor{lightblack}22\%  & --  &  \cellcolor{lightblue}7\%   & --  & -- \\
\multicolumn{1}{l}{LIC} & \cellcolor{lightblack}17\%  & \cellcolor{lightgreen}84\%  &  \cellcolor{lightblue}8\%   & \cellcolor{lightgreen}67\%  &  \cellcolor{lightblue}7\%   & \cellcolor{lightgreen}73\%  & \cellcolor{lightblack}19\%  & \cellcolor{lightgreen}89\%  &  \cellcolor{lightblack}15\%   & \cellcolor{lightgreen}81\%  &  \cellcolor{lightblack}19\%   & \cellcolor{lightgreen}88\% \\
\multicolumn{1}{l}{USA} & \cellcolor{lightblack}36\%  & \cellcolor{lightgreen}92\%  & \cellcolor{lightblack}34\%  & \cellcolor{lightgreen}72\%  &  \cellcolor{lightblue}9\%   & \cellcolor{lightgreen}87\%  & \cellcolor{lightblack}22\%  & --  & \cellcolor{lightblack}17\%  & --  &  \cellcolor{lightblue}9\%   & -- \\
\multicolumn{1}{l}{VFR} & \cellcolor{lightblack}20\%  & \cellcolor{lightblack}41\%  & \cellcolor{lightblack}18\%  & \cellcolor{lightgreen}51\%  & \cellcolor{lightblack}15\%  & \cellcolor{lightgreen}62\%  & \cellcolor{lightblack}15\%  & \cellcolor{lightblack}40\%  & \cellcolor{lightblue}10\%  & \cellcolor{lightblack}44\%  & \cellcolor{lightblue}9\%  & \cellcolor{lightgreen}53\% \\
\multicolumn{1}{l}{MWN} & \cellcolor{lightblack}21\%  & \cellcolor{lightblack}35\%  & \cellcolor{lightblack}16\%  & \cellcolor{lightblack}35\%  & \cellcolor{lightblack}22\%  & \cellcolor{lightblack}49\%  & \cellcolor{lightblack}17\%  & \cellcolor{lightblack}31\%  & \cellcolor{lightblack}12\%  & \cellcolor{lightblack}33\%  & \cellcolor{lightblack}11\%  & \cellcolor{lightblack}46\% \\
\multicolumn{1}{l}{AE} & \cellcolor{lightblack}40\%  &  \cellcolor{lightblack}14\%   & \cellcolor{lightblack}39\%  & \cellcolor{lightgreen}73\%  &  \cellcolor{lightblack}14\%   & \cellcolor{lightgreen}87\%  & \cellcolor{lightblack}16\%  &  --   & \cellcolor{lightblack}16\%  & --  &  \cellcolor{lightblack}11\%   & -- \\
\multicolumn{1}{l}{DOM} & \cellcolor{lightblack}54\%  & \cellcolor{lightgreen}97\%  & \cellcolor{lightblack}48\%  & \cellcolor{lightgreen}70\%  &  \cellcolor{lightblue}6\%   & \cellcolor{lightgreen}95\%  & \cellcolor{lightblack}27\%  & --  & \cellcolor{lightblack}32\%  & --  &  \cellcolor{lightblue}6\%   & -- \\
\multicolumn{1}{l}{LMA} & \cellcolor{lightblack}35\%  & \cellcolor{lightblack}46\%  & \cellcolor{lightblack}28\%  & \cellcolor{lightgreen}74\%  &  \cellcolor{lightblue}6\%   & \cellcolor{lightblack}46\%  & \cellcolor{lightblack}22\%  & --  & \cellcolor{lightblack}13\%  & --  &  \cellcolor{lightblue}6\%   & -- \\
\multicolumn{1}{l}{LMNA} & \cellcolor{lightblack}29\%  & \cellcolor{lightblack}39\%  & \cellcolor{lightblack}22\%  & \cellcolor{lightgreen}68\%  &  \cellcolor{lightblue}9\%   & \cellcolor{lightblack}41\%  & \cellcolor{lightblack}21\%  & --  & \cellcolor{lightblack}15\%  & --  &  \cellcolor{lightblue}7\%   & -- \\
\multicolumn{1}{l}{LV} & \cellcolor{lightblack}38\%  & \cellcolor{lightblack}30\%  & \cellcolor{lightgreen}84\%  &  \cellcolor{lightblack}11\%   & \cellcolor{lightblack}56\%  & \cellcolor{lightgreen}75\%  & \cellcolor{lightblack}25\%  & \cellcolor{lightblack}21\%  & \cellcolor{lightgreen}68\%  &  \cellcolor{lightblack}11\%   & \cellcolor{lightblack}69\%  & \cellcolor{lightgreen}80\% \\
\multicolumn{1}{l}{NA} & \cellcolor{lightblack}26\%  & \cellcolor{lightgreen}88\%  & \cellcolor{lightblack}30\%  & \cellcolor{lightblack}50\%  &  \cellcolor{lightblue}10\%   & \cellcolor{lightgreen}80\%  & \cellcolor{lightblack}13\%  & \cellcolor{lightgreen}87\%  & \cellcolor{lightblack}12\%  & \cellcolor{lightgreen}71\%  &  \cellcolor{lightblue}10\%   & \cellcolor{lightgreen}85\% \\
\multicolumn{1}{l}{PDOM} & \cellcolor{lightblack}51\%  & \cellcolor{lightblack}41\%  & \cellcolor{lightgreen}95\%  &  \cellcolor{lightblue}10\%   & \cellcolor{lightgreen}72\%  & \cellcolor{lightgreen}95\%  & \cellcolor{lightblack}24\%  & \cellcolor{lightblack}20\%  & --  &  \cellcolor{lightblack}24\%   & --  & -- \\
\multicolumn{1}{l}{RD} & \cellcolor{lightblack}15\%  & \cellcolor{lightgreen}80\%  &  \cellcolor{lightblue}7\%   & \cellcolor{lightgreen}62\%  &  \cellcolor{lightblue}9\%   & \cellcolor{lightgreen}68\%  & \cellcolor{lightblack}19\%  & \cellcolor{lightgreen}91\%  &  \cellcolor{lightblue}10\%   & \cellcolor{lightgreen}79\%  &  \cellcolor{lightblue}5\%   & \cellcolor{lightgreen}86\% \\
\multicolumn{1}{l}{RS} & \cellcolor{lightblack}31\%  & \cellcolor{lightblack}31\%  & \cellcolor{lightblack}30\%  & \cellcolor{lightblack}31\%  & \cellcolor{lightblack}28\%  & \cellcolor{lightblack}30\%  & \cellcolor{lightblack}16\%  & \cellcolor{lightblack}40\%  & \cellcolor{lightblue}9\%  & \cellcolor{lightblack}31\%  & \cellcolor{lightblue}7\%  & \cellcolor{lightblack}49\% \\
\multicolumn{1}{l}{VBE} & \cellcolor{lightblack}40\%  & \cellcolor{lightblack}36\%  & \cellcolor{lightgreen}88\%  &  \cellcolor{lightblack}13\%   & \cellcolor{lightgreen}76\%  & \cellcolor{lightgreen}81\%  & \cellcolor{lightblack}28\%  & \cellcolor{lightblack}24\%  & --  &  \cellcolor{lightblue}10\%   & --  & -- \\
\multicolumn{1}{l}{SS} & \cellcolor{lightblack}26\%  & \cellcolor{lightblack}39\%  & \cellcolor{lightblack}22\%  & \cellcolor{lightblack}37\%  & \cellcolor{lightblack}25\%  & \cellcolor{lightgreen}57\%  & \cellcolor{lightblack}20\%  & \cellcolor{lightblack}35\%  & \cellcolor{lightblack}13\%  & \cellcolor{lightblack}34\%  & \cellcolor{lightblue}10\%  & \cellcolor{lightgreen}50\% \\
\multicolumn{1}{l}{UDV} &  \cellcolor{lightblue}6\%   &  \cellcolor{lightblue}5\%   &  \cellcolor{lightblue}6\%   &  \cellcolor{lightblue}10\%   &  \cellcolor{lightblue}9\%   &  \cellcolor{lightblue}3\%   &  \cellcolor{lightblue}3\%   &  \cellcolor{lightblue}4\%   &  \cellcolor{lightblue}2\%   &  \cellcolor{lightblue}7\%   &  \cellcolor{lightblue}6\%   &  \cellcolor{lightblue}0\% \\
\multicolumn{1}{l}{UIR} &  \cellcolor{lightblue}2\%   &  \cellcolor{lightblue}2\%   &  \cellcolor{lightblue}1\%   & \cellcolor{lightblue}3\%   & \cellcolor{lightblue}3\%   & \cellcolor{lightblue}0\%   &  \cellcolor{lightblue}2\%   &  \cellcolor{lightblue}5\%   &  \cellcolor{lightblue}4\%   & \cellcolor{lightblue}7\%   & \cellcolor{lightblue}7\%   & \cellcolor{lightblue}0\% \\
\multicolumn{1}{l}{WNIL} &  \cellcolor{lightblue}3\%   &  \cellcolor{lightblue}4\%   &  \cellcolor{lightblue}5\%   &  \cellcolor{lightblue}6\%   &  \cellcolor{lightblue}8\%   & \cellcolor{lightblue}2\%  &  \cellcolor{lightblue}3\%   &  \cellcolor{lightblue}6\%   &  \cellcolor{lightblue}5\%   &  \cellcolor{lightblue}5\%   &  \cellcolor{lightblue}6\%   & \cellcolor{lightblue}0\% \\
\midrule
\multicolumn{1}{l}{Overall} &  \cellcolor{lightblack}31\%   &  \cellcolor{lightgreen}83\%   &  \cellcolor{lightgreen}70\%   &  \cellcolor{lightblack}55\%   &  \cellcolor{lightblack}35\%   &  \cellcolor{lightgreen}81\%   &  --   &  --  &  --   &  --   &  --  &  -- \\
\bottomrule
\end{tabular}%
}

\subfloat[Reduction over analysis
properties.\label{tab:reduction-analysis-properties}] {
\begin{tabular}{lrrrrrr}
\toprule
\multicolumn{1}{l}{Property} & \multicolumn{6}{c}{DaCapo}   \\
\cmidrule(lr){2-7}
      & DFS   & PO    & RPO   & WPO   & WRPO  & ANY   \\
\midrule
\multicolumn{1}{l}{Data-flow} & \cellcolor{lightblack}32\%  & \cellcolor{lightgreen}84\%  &  \cellcolor{lightgreen}72\%   & \cellcolor{lightgreen}57\%  &  \cellcolor{lightblack}36\%   & \cellcolor{lightgreen}83\% \\
\multicolumn{1}{l}{$\neg$Data-flow} & \cellcolor{lightblue}4\%  & \cellcolor{lightblue}4\%  & \cellcolor{lightblue}4\%  & \cellcolor{lightblue}6\%  &  \cellcolor{lightblue}6\%   & \cellcolor{lightblue}2\% \\
\bottomrule
\end{tabular}%
}
\subfloat[Reduction over graph
properties.\label{tab:reduction-graph-properties}] {
\begin{tabular}{lrrrrrr}
\toprule
\multicolumn{1}{l}{Property} & \multicolumn{6}{c}{DaCapo} \\
\cmidrule(lr){2-7} 
      & DFS   & PO    & RPO   & WPO   & WRPO  & ANY   \\
\midrule
\multicolumn{1}{l}{Sequential} & \cellcolor{lightblack}20\%  & \cellcolor{lightgreen}74\%  &  \cellcolor{lightgreen}63\%   & \cellcolor{lightgreen}55\%  &  \cellcolor{lightblack}28\%   & \cellcolor{lightgreen}72\%  \\
\multicolumn{1}{l}{Branch} & \cellcolor{lightblack}31\%  & \cellcolor{lightgreen}81\%  & \cellcolor{lightgreen}66\%  & \cellcolor{lightgreen}58\%  &  \cellcolor{lightblack}40\%   & \cellcolor{lightgreen}92\% \\
\multicolumn{1}{l}{Loop} & \cellcolor{lightgreen}53\%  & \cellcolor{lightgreen}88\%  & \cellcolor{lightgreen}75\%  &  \cellcolor{lightgreen}62\%   & \cellcolor{lightblack}37\%  & \cellcolor{lightgreen}95\% \\
\bottomrule
\end{tabular}%
}
\caption{Reduction in running times. \footnotesize Background colors indicate ranges of 
values: \colorbox{lightred}{no reduction}, \colorbox{lightblue}{(0\%, 10\%)}, 
\colorbox{lightblack}{[10\%, 50\%)} and \colorbox{lightgreen}{[50\%, 100\%]}.}
\label{fig:reduction}
\end{figure*}

To evaluate the efficiency in running time of \shortname over other 
strategies, we ran 21 analyses on DaCapo and GitHub 
datasets using \shortname and other strategies. When comparing 
the \shortname to a standard strategy $S$, we computed the reduction 
rate $R=(T_S-T_H)/T_S$ where $T_S$ and $T_H$ are the running times using the 
standard strategy and \shortname, respectively. 
Some analyses have worst case traversal strategies which might not be 
feasible to run on GitHub dataset with 162 million graphs. 
For example, using post-order for forward data-flow 
analysis will visit the CFGs in the direction which is opposite to the 
natural direction of the analysis and hence takes a long time to complete. 
For such combinations of analyses and traversal strategies, the 
map and the reduce tasks time out in the cluster setting and, thus, did 
not have the running times. 
The corresponding cells in \figref{tab:reduction-analyses} are denoted with symbol --.
%

The result in \figref{tab:reduction-analyses} shows that \shortname 
helps reduce the running times in almost all cases. The values indicate the 
reduction in running time by adopting \shortname compared against the standard strategies. 
Most of positive reductions are from \colorbox{lightblack}{10\%} or even from 
\colorbox{lightgreen}{50\%}. 
Compared to the most time-efficient strategies for each analysis, \shortname could 
speed up from 1\% (UIR with RPO) to 28\% (RS with WRPO).
More importantly, the most time-efficient and the worst traversal strategies vary 
across the analyses which supports the need of \shortname.
Over all analyses, the reduction was highest against any order and 
post-order (PO and WPO) strategies. The reduction was lowest against the strategy 
using depth-first search (DFS) and worklist with reverse post-ordering (WRPO). 
When compared with the next best performing traversal strategy for each analysis,  \shortname reduces the overall execution time by 
about 13 minutes to 72 minutes on GitHub dataset. We do not 
report the overall numbers for GitHub dataset due to the presence of failed 
runs.

\figref{tab:reduction-analysis-properties} shows time reductions for 
different types of analyses. For \textit{data-flow sensitive} ones, the 
reduction rates were high ranging from 32\% to 84\%. 
The running time was not improved much for \emph{non data-flow sensitive} 
traversals, which correspond to the last three rows in \figref{tab:reduction-analyses} 
with mostly \colorbox{lightblue}{one digit reductions}). 
We actually perform almost the same as ANY-order traversal 
strategy for analyses in this category. This is because any-order traversal 
strategy is the best strategy for all the CFGs in these analyses.
\shortname also chooses any-order traversal strategy and, thus, 
the performance is the same.

\figref{tab:reduction-graph-properties} shows time reduction for
different \graphprop{} types of input graphs. We can see that
reductions over graphs with loops is highest and those over any graphs
is lowest.

\subsection{Correctness of Analysis Results}
\label{sec:soundness}

To evaluate the correctness of analysis results, we first chose worklist 
as standard strategy to run analyses on DaCapo dataset to create the 
groundtruth of the results. We then ran analyses using our hybrid approach 
and compared the results with the groundtruth. In all analyses on all 
input graphs from the dataset, the results from \shortname always 
exactly matched the corresponding ones in the groundtruth. 

\subsection{Traversal Strategy Selection Precision}
\label{sec:prediction-accuracy}

In this experiment, we evaluated how well \shortname picks the most 
time-efficient strategy. We ran the 21 analyses on the DaCapo dataset using 
all the candidate traversals and the one selected by \shortname. One 
selection is counted for each pair of a traversal and an input graph where 
the \shortname selects a traversal strategy based on the properties of 
the analysis and input graph. A selection is considered correct if its 
running time is at least as good as the running time of the fastest among all 
candidates. The precision is computed as the ratio between the number of 
correct selections over the total number of all selections. The precision was 
100\% and 99.9\% for \textit{loop insensitive} and \textit{loop sensitive} traversals, respectively.

\begin{table}
\footnotesize
\centering
\caption{Traversal strategy prediction precision.}
\setlength{\tabcolsep}{0.5pt}
\begin{tabular}{cr}
\toprule
Analysis & Precision \\
\midrule
DOM, PDOM, WNIL, UDV, UIR & 100.00\% \\
CP, CSD, DC, LIC, USA, VFR, MWN, AE, LMA, LMNA, LV, NA, RD, RS, VBE, SS  & 99.99\% \\
\bottomrule
\end{tabular}%
\label{tab:prediction-precision}%
\end{table}%
As shown in \tabref{tab:prediction-precision}, the selection precision is 100\%
for all analyses that are not \textit{loop sensitive}. For analyses that
involve \textit{loop sensitive} traversals, the prediction precision is 99.99\%.
Further analysis revealed that the selection precision is 100\% for 
sequential CFGs \& CFGs with branches and no loop---\shortname always
picks the most time-efficient traversal strategy.
For CFGs with loops, the selection precision is 100\% for \emph{loop 
insensitive} traversals. The mispredictions occur with \emph{loop sensitive} 
traversals on CFGs with loops.
This is because for \textit{loop sensitive} traversals, \shortname 
picks worklist as the best strategy. The worklist approach was 
picked because it visits only as many nodes as needed when compared to other 
traversal strategies which visit redundant nodes. However using worklist 
imposes an overhead of creating and maintaining a worklist containing all 
nodes in the CFG. This overhead is negligible for small CFGs. However, when 
running analyses on large CFGs, this overhead could become higher 
than the cost for visiting redundant nodes. Therefore, selecting worklist for 
\emph{loop sensitive} traversals on large CFGs might not always result in the 
best running times. 

\subsection{Analysis on Traversal Optimization}
\label{sec:analysis-traversal-opt}

\begin{figure}
\centering
\includegraphics[width=\linewidth]{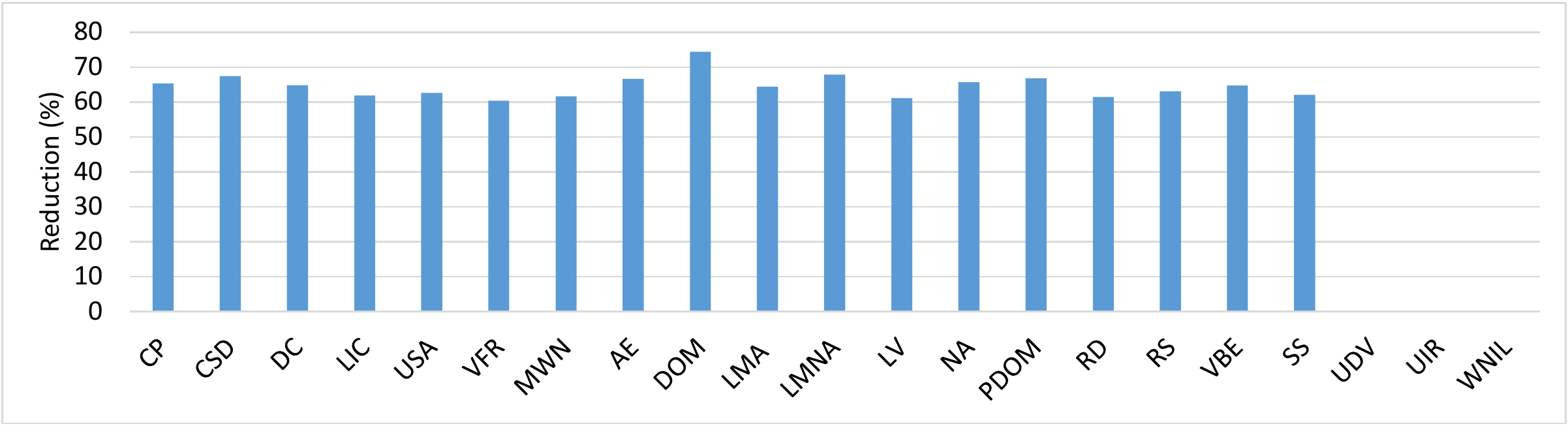}
\caption{Time reduction due to traversal optimization.}%
\label{fig:hy-opt-off}%
\end{figure}

We evaluated the importance of optimizing the chosen traversal strategy by 
comparing \shortname with the non-optimized version. 
\fignref{fig:hy-opt-off} shows the reduction rate on the running times for the 21 analyses. 
For analyses that involve at least one 
\textit{data-flow sensitive} traversal, the optimization helps to reduce at 
least 60\% of running time. This is because optimizations in such traversals 
reduce the number of iterations of traversals over the graphs by eliminating 
the redundant result re-computation traversals and the
unnecessary fixpoint condition checking traversals. For analyses involving only 
\textit{data-flow insensitive} traversal, there is no reduction in execution time, as 
\shortname does not attempt to optimize.

%
%
\subsection{Case Studies}
\label{sec:case-studies}
This section presents three applications adopted from prior works that showed
significant benefit from \shortname approach. These
applications includes one or more analyses listed in
\tabref{tab:analysis-table}. We computed the reduction in the overall analysis
time when compared to WRPO traversal strategy (the second best performing
traversal after \shortname) and the results are shown in
\autoref{fig:case-study-table}.

%

\begin{figure}[ht!]%
\centering
\footnotesize
\begin{tabular}{lrrr}
\toprule
Case & \shortname & WRPO & Reduce \\
\midrule
APM & 1527 min. & 1702 min. & 10\% \\
AUM & 883 min. & 963 min. & 8\% \\
SVT & 1417 min. & 1501 min. & 6\% \\
\bottomrule
\end{tabular}%
\caption{Running time of the case studies on GitHub data.}
\label{fig:case-study-table}
\end{figure}

\textbf{API Precondition Mining (APM).} This case study mines a large corpus 
of API usages to derive potential preconditions for API methods~\cite{preconditions}. 
The key idea is that API preconditions would be checked frequently in a 
corpus with a large number of API usages, while project-specific 
conditions would be less frequent. 
This case study mined the preconditions for all methods of 
\lstinline|java.lang.String|. 

\textbf{API Usage Mining (AUM).} This case study analyzes API usage code and 
mines API usage patterns~\cite{xie2006mapo}. The mined patterns help 
developers understand and write API usages more effectively 
with less errors. Our analysis mined usage patterns for 
\lstinline|java.util| APIs. 

\textbf{Finding Security Vulnerabilities with Tainted Object 
Propagation (SVT).} This case study formulated a variety of widespread SQL 
injections, as tainted object propagation problems~\cite{livshits2005finding}. 
Our analysis looked for all SQL injection vulnerabilities 
matching the specifications in the statically analyzed code. 

\autoref{fig:case-study-table} shows that \shortname helps reduce running times
significantly by 80--175 minutes, which is from 6\%--10\% relatively. For
understanding whether ~10\% reduction is really significant, considering the
context is important. A save of 3 hours (10\%) on a parallel infrastructure is
significant. If the underlying parallel infrastructure is open/free/shared
(\cite{dyer2013boa,Bajracharya-etal-04}), a 3 hour save enables supporting more
concurrent users and analyses. If the infrastructure is a paid cluster
(e.g., AWS), a 3 hour less computing time could translate to save of substantial
dollar amount.

%

\subsection{Threats to Validity}
\label{sec:threats}

Our datasets do not contain a balanced
distribution of different graph \graphprop{}.
The majority of graphs in both DaCapo and GitHub datasets are sequential (65\% and
69\%, respectively) and only 10\% have loops. The impact of this
threat is that paths and decisions along sequential graphs are taken more
often. This threat is not easy to mitigate, as it is not pratical to find
a code dataset with a balanced distribution of graphs
of various types. Nonetheless, our evaluation shows that the selection and
optimization of the best traversal strategy for these 35\% of the graphs (graphs
with branches and loops) plays an important role in improving the overall
performance of the analysis over a large dataset of graphs.
\section{Related Works}
\label{sec:related}

Atkinson and Griswold~\cite{atkinson2001implementation} discuss several
implementation techniques for improving the efficiency of data-flow analysis,
namely: factoring data-flow sets, visitation order of the statements, selective
reclamation of the data-flow sets. They discuss two commonly used traversal
strategies: iterative search and worklist, and propose a new worklist algorithm
that results in 20\% fewer node visits. In their algorithm, a node is processed
only if the data-flow information of any of its successors (or predecessors) has
changed.
Tok\etal~\cite{tok-etal} proposed a new worklist algorithm for accelerating
inter-procedural flow-sensitive data-flow analysis. They generate
inter-procedural def-use chains on-the-fly to be used in their worklist
algorithm to re-analyze only parts that are affected by the changes in the flow
values.
Hind and Pioli~\cite{hind-pioli} proposed an optimized priority-based worklist
algorithm for pointer alias analysis, in which the nodes awaiting processing
are placed on a worklist prioritized by the topological order of the CFG, such
that nodes higher in the CFG are processed before nodes lower in the CFG.
Bourdoncle~\cite{bourdoncle93} proposed the notion of weak topological ordering
(WTO) of directed graphs and two iterative strategies based on WTO for computing
the analysis solutions in dataflow and abstraction interpretation domains.
Bourdoncle' technique is more suitable for cyclic graphs, however for acyclic
graphs Bourdoncle proposes any topological ordering.
Kildall~\cite{kildall1973unified} proposes combining several optimizing
functions with flow analysis algorithms for solving global code optimization
problems. For some classes of data-flow analysis problems, there exist
techniques for efficient analysis. For example, demand interprocedural data-flow
analysis~\cite{horwitz-etal} can produce precise results in polynomial time for
inter-procedural, finite, distributive, subset problems (IFDS), constant
propagation~\cite{wegman-etal}, etc.
These works propose new traversal strategies for improving the efficiency of
certain class of source code analysis, whereas \shortname is a novel technique for selecting the best
traversal strategy from a list of candidate traversal strategies, based on the
static properties of the analysis and the runtime characteristics of the input
graph.

Upadhyaya and Rajan~\cite{Upadhyaya2018} proposed Collective Program Analysis
(CPA) that leverages similarities between CFGs to speedup analyzing millions of
CFGs by only analyzing unique CFGs. CPA utilizes pre-defined traversal
strategy to traverse CFGs, however our technique selects optimal traversal
strategy and could be utilized in CPA.
Upadhyaya and Rajan have also proposed an approach for accelerating ultra-large
scale mining by clustering artifacts that are being mined~\cite{upadhyaya2017accelerating,Upadhyaya2018b}.
BCFA and this approach have the same goal of scaling large-scale mining, 
but complementary strategies.

Cobleigh\etal~\cite{cobleigh-etal} study the effect of worklist algorithms in
model checking. They identified four dimensions along which a worklist algorithm
can be varied. Based on four dimensions, they evaluate 9 variations of worklist
algorithm. They do not solve traversal strategy selection problem. Moreover,
they do not take analysis properties into account. 
We consider both static properties of the analysis, such as data-flow
sensitivity and loop sensitivity, and the cyclicity of the graph. Further, we
also consider non-worklist based algorithms, such as post-order, reverse
post-order, control flow order, any order, etc., as candidate strategies.

Several infrastructures exist today for performing ultra-large-scale
analysis~\cite{dyer2013boa,Bajracharya-etal-04,Gousi13,heritage,world-of-code}.
Boa~\cite{dyer2013boa} is a language and infrastructure for analyzing open source projects. 
Sourcerer~\cite{Bajracharya-etal-04} is an infrastructure for
large-scale collection and analysis of open source code.
GHTorrent~\cite{Gousi13} is a dataset and tool suite for analyzing GitHub
projects. These frameworks currently support structural or abstract syntax tree
(AST) level analysis and a parallel framework such as map-reduce is used to
improve the performance of ultra-large-scale analysis. 
By selecting the best traversal strategy, \shortname could help improve their performance
beyond parallelization.

There have been much works that targeted graph traversal optimization. 
Green-Marl~\cite{hong2012green} is a domain specific language for
expressing graph analysis. 
It uses the high-level algorithmic description of the
graph analysis to exploit the exposed data level
parallelism.
Green-Marl's optimization is similar to ours in
utilizing the properties of the analysis description, however \shortname also
utilizes the properties of the graphs.
Moreover, Green-Marl's optimization is through parallelism while ours is by selecting
the suitable traversal strategy.
Pregel~\cite{malewicz2010pregel} is a map-reduce like framework that aims to
bring distributed processing to graph algorithms. While Pregel's performance gain
is through parallelism, \shortname achieves it by traversing the
graph efficiently.

\section{Conclusion and Future Work}
\label{conclusion}
Improving the performance of source code analyses that runs on massive code bases
is an ongoing challenge. 
We proposed \name, a technique for optimizing source code
analysis over control flow graphs (CFGs).
Given the code of the analysis and a large set of CFGs, \shortname extracts a
set of static properties of the analysis and combines it with a property of the
CFG to automatically select an optimal CFG traversal strategy for every
CFG. \shortname has minimal overhead, less than 0.2\%; and leads to speedup
between 1\%-28\%. \shortname has been integrated into Boa~\cite{dyer2013boa,boa-b} and
has already been utilized in ~\cite{zhang-etal}.


An immediate avenue for future work lies in extending our technique to analyses
that go beyond method boundaries (program-level), as our current work only
targets method-level source code analysis. We also plan to extend our technique
to other source code graphs that requires traversing, for instance, program
dependence graphs (PDGs) and call graphs (CGs).
Boa has been used for studies that require analyses~\cite{Maddox18,zhang-etal} and we would like 
to understand if those analyses can benefit from this work. 


\section*{Acknowledgments}
This work was supported in part by US NSF under grants CNS-15-18897, and
CNS-19-34884. All opinions are of the authors and do not reflect the view of
sponsors. We thank ICSE'20 reviewers for constructive comments that were very
helpful.


\bibliographystyle{ACM-Reference-Format}
\bibliography{refs}
\end{document}